\documentclass[aps, reprint,10.0pt,twocolumn,superscriptaddress]{revtex4-1}
\usepackage{amsmath}
\usepackage{latexsym}
\usepackage{amssymb}
\usepackage{graphics,epstopdf}
\usepackage{hyperref}
\usepackage{soul}
\hypersetup{
    colorlinks = true,
    linkcolor =blue,
	citecolor=blue, 
	urlcolor=blue 
}

\newcommand{\ed}{\end{document}}
\newcommand{\beq}{\begin{equation}}
\newcommand{\eeq}{\end{equation}}

\usepackage{mathtools}
\usepackage{natbib}

\begin{document}
\title{Environment assisted superballistic scaling of conductance}

\author{Madhumita Saha}
\email{madhumita.saha@acads.iiserpune.ac.in} 
\affiliation{Department of Physics, Indian Institute of Science Education and Research Pune, Dr. Homi Bhabha Road, Ward No. 8, NCL Colony, Pashan, Pune, Maharashtra 411008, India}
\affiliation{International Centre for Theoretical Sciences, Tata Institute of Fundamental Research,
Bangalore 560089, India}

\author{Bijay Kumar Agarwalla}
\email{bijay@iiserpune.ac.in}
\affiliation{Department of Physics, Indian Institute of Science Education and Research Pune, Dr. Homi Bhabha Road, Ward No. 8, NCL Colony, Pashan, Pune, Maharashtra 411008, India}

\author{Manas Kulkarni}
\email{manas.kulkarni@icts.res.in} 
\affiliation{International Centre for Theoretical Sciences, Tata Institute of Fundamental Research,
Bangalore 560089, India}

\author{Archak Purkayastha}
\email{archak.p@phy.iith.ac.in}
\affiliation{School of Physics, Trinity College Dublin, College Green, Dublin 2, Ireland}
\affiliation{Centre for complex quantum systems, Aarhus University, Nordre Ringgade 1, 8000 Aarhus C, Denmark}
\affiliation{Department of Physics, Indian Institute of Technology, Hyderabad 502284, India}

\date{\today} 
\begin{abstract}
{We find that, in the presence of weak incoherent effects from surrounding environments, the low temperature conductance of nearest neighbour tight-binding fermionic chains exhibits a counter-intuitive monotonic growth with system length when Fermi energy is near the band edges, indicating a superballistic scaling. This fascinating environment assisted superballistic scaling of conductance occurs over a finite but extended regime of system lengths. This regime can be systematically expanded by decreasing the coupling to the surrounding environments and by reducing temperature. This behavior is robust against weak disorder and slight shifts from band edge, although the extent of the superballistic scaling regime is affected by them. We give precise predictions of how the superballistic scaling regime depends on coupling to surrounding environments, disorder strength, shifts from band edge and temperature. There is no corresponding analog of this behavior in isolated systems. The superballistic scaling stems from an intricate interplay of incoherent effects from surrounding environments and exceptional points of the system's transfer matrix that occur at every band edge.
}
\end{abstract}

\maketitle

{\it Introduction --} 
The resistance of a normal metal wire is proportional to its length, indicating diffusive transport. As a result, the metal's resistivity, given by resistance per unit length per unit cross-sectional area, is well-defined. Deviation from this diffusive behavior, which leads to ill-defined resistivity, can be seen in a variety of situations, particularly in low-dimensional systems, and has been of great research interest \cite{length_dependent,length_thermal_molecular_chain,divergent_thermal_conductivity_graphene,divergent_thermal_conductivity,size_thermal_graphene,Manas-anomalous-1, Manas-anomalous-2,Xu_2016, Dhar_2019, Dhar_2008, Bertini_2021, Landi_2022}. Even outside of the diffusive regime, resistance generically increases with system length. The main exception is that of perfectly ballistic transport where resistance does not scale with system length \cite{Li_2020,Homoth_2009,Liu_2005,Li_2005,Debray_2000}.

In this letter, we demonstrate the possibility of behavior different from all of the above: resistance of a wire can decay monotonically over a finite but large regime of system lengths. In other words, there exists a regime in which conductance, i.e. the inverse of resistance, can increase monotonically with system length, thereby exhibiting superballistic scaling. This rather counter-intuitive behavior occurs close to zero temperature near the band edges of the system, assisted by weak incoherent effects from the surrounding environments. The regime exhibiting superballistic scaling systematically expands on weakening the system's coupling to its surrounding environments without completely isolating it from them.

We find this intriguing behavior by combining concepts from non-Hermitian physics \cite{Feng2017,El_Ganainy2018,El_Ganainy2019,Ashida_2020} with those from quantum chemistry and mesoscopic physics. Borrowing from the latter, we model the surrounding environments by B\"uttiker voltage probes (BVPs) \cite{buttiker_paper,molecular_buttiker,molecular_dephasing1, molecular_dephasing2, molecular_dephasing4,segal-probe-1,segal-probe-2,segal-probe-3,malay,Bedkihal2013,Pastawski, Roy-1,self-2-Abhishek, dephasing_dubi,kulkarni2013full,kulkarni2012towards}. We show that the superballistic scaling of conductance near every band edge arises from an interplay of the incoherent effects from the BVPs and exceptional points (EPs) of the system's transfer matrix that occurs at every band edge \cite{saha2022universal}. The transfer matrix is a non-Hermitian matrix that appears in scattering theory. It plays a fundamental role in determining the band structure of the system and its transport properties \cite{lastY, molinari1, molinari2}. To our knowledge,  the role of non-Hermitian properties of the transfer matrix on environment assisted transport has remained completely unexplored, despite the later being investigated both theoretically and experimentally, across physics, chemistry and biology \cite{Plenio_2008,Rebentrost_2009,Caruso_2009, Sowa_2017,Harush_2018,Dutta_2017,buttikerdephase,segal-probe-1,segal-probe-2,segal-probe-3,dephasing_dubi,Landi-latest,ENQT1,diffusion1,buttikerdephase, saha_buttiker, dephasing-interaction1, dephasing_spin_interaction2, dephasing_interaction3, Mendoza-Arenas_2013,Maier_2019,Leon-Montiel_2015,Biggerstaff_2016,Viciani_2015,Harris_2017}.

It is worth mentioning that the term `superballistic' has been used in various separate contexts. In some experiments, conductance larger than the maximum conductance of free electrons has been termed superballistic \cite{superballistic1,superballistic2,superballistic3}. In a separate set of works, faster-than-ballistic spread of an initially localized wavepacket has been explored both theoretically and experimentally \cite{superballistic4, Superballistic5,Superballistic6}. However, to our knowledge, the superballistic scaling of conductance with system length has not been reported before. Unlike the spread of an initially localized wavepacket, this feature crucially requires presence of incoherent effects from surrounding environments and therefore cannot be seen in an isolated system.

{\it  Lattice chain with BVPs--}  We consider a nearest neighbour tight-binding lattice chain consisting of $N$ sites. For simplicity, we consider the single-band Hamiltonian given by $\hat{H}_C= \sum_{n=1}^{N-1} g\left(\hat{c}^{\dagger}_n \hat{c}_{n+1} + \hat{c}^{\dagger}_{n+1} \hat{c}_{n}\right)+\varepsilon \hat{H}_{\rm dis}$, where $\hat{c}_{n}$ is the fermionic annihilation operator at the $n$-th site. The Hamiltonian $\hat{H}_{\rm dis}$ contains random quadratic disorder terms in on-site energies and hopping. The parameter $\varepsilon$ controls the overall strength of disorder.  This chain is attached to a source bath with chemical potential $\mu_L$ at the left end, i.e, at first site, and a drain bath with chemical potential $\mu_R$ at the right end, i.e, at $N$-th site, which drives a current through the chain.  Each bath is modelled via an infinite number of fermionic modes, and the system bath couplings are taken bilinear and number conserving. For simplicity, we consider the wide-band limit, with $\tau$ giving the effective strength of coupling with the left and right baths. The source and drain can be arbitrarily strongly coupled, we do not put any restriction on the magnitude of $\tau$.

\begin{figure}
\includegraphics[width=\columnwidth]{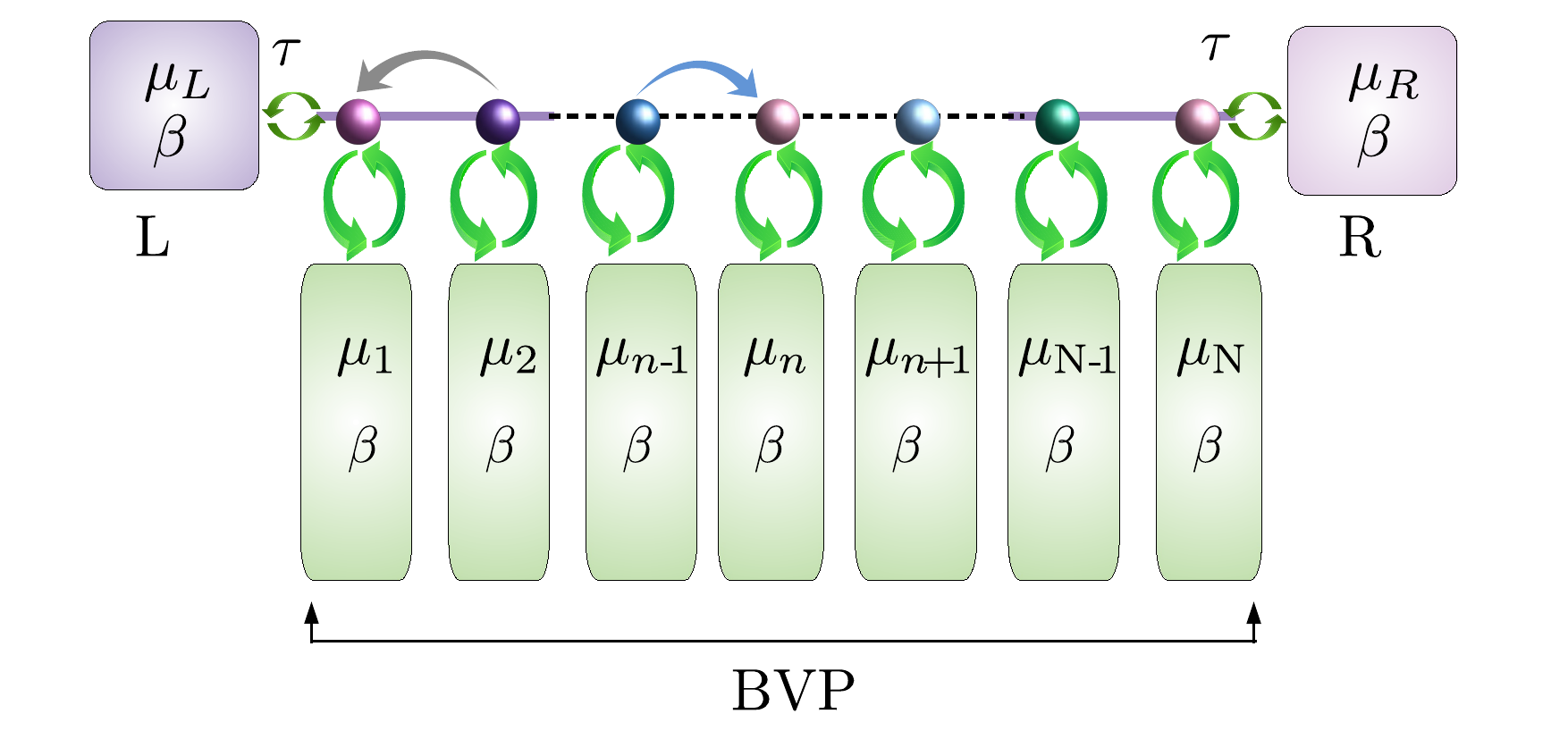} 
\caption{A schematic of our set-up showing a tight-binding chain subjected to left bath (L), right bath (R) and BVPs, all of which are at the same inverse temperature $\beta$. The left and right baths are coupled with strength $\tau$, while the $n$th BVP is coupled with strength $\tau_P\nu^{(n)}$. The chemical potential of the probes $\mu_n, n=1,2, \cdots N$ are determined by demanding the zero particle current between $n$-th site and $n$-th probe.} 
\label{schematic} 
\end{figure}

This lattice chain is subject to weak incoherent effects from surrounding environments apart from the source-drain baths. This is modelled by attaching BVPs at all the sites. These are baths similar to the source and drain baths, except that their chemical potentials, $\{\mu_n\}$, are such that there is no average particle current into each of them. The temperature of all the baths are considered same, given by inverse temperature $\beta$. The microscopic Hamiltonian $\hat{H}$ for this whole set-up can be written as, $\hat{H}=\hat{H}_C + \hat{H}_L + \hat{H}_R + \sum^\prime \hat{H}_{P_n}+\hat{H}_{CL} + \hat{H}_{CR} + \sum^\prime \hat{H}_{CP}^{(n)}$, $\hat{H}_L$ is the Hamiltonian of the left bath (source), $\hat{H}_R$ is the Hamiltonian of the right bath (drain), $\hat{H}_{P_n}$ is the Hamiltonian of the probe attached to the $n$-th site of the system, $\hat{H}_{CL}$, $\hat{H}_{CR}$ and $\hat{H}_{CP}^{(n)}$ are the Hamiltonians which describe the coupling of central system with source, drain, and $n$-th probe respectively, and $\sum^\prime$ denotes sum over the sites where the BVPs are attached.  
A schematic of our entire set-up is shown in Fig.~(\ref{schematic}). We describe the set-up in terms of the retarded non-equilibrium Green's function (NEGF) of the system, given by 
$
G(\omega)=\left[\omega \mathbb{I}\!-\!H_C\!-\!\Sigma_L(\omega)-\Sigma_R(\omega)\!-\!\sum_{n=1}^{N} \Sigma_{P_n}(\omega)\right]^{-1}
$. Here $\mathbb{I}$ is the $N \times N$ identity matrix, $H_C$ is the $N \times N$ single particle Hamiltonian corresponding to $\hat{H}_C$, and $\Sigma_L(\omega)$, $\Sigma_R(\omega)$, $\Sigma_{P_n}(\omega)$ are the retarded self-energy matrices of the left, right and probe baths respectively, the only non-zero elements of which are $[\Sigma_{L}(\omega)]_{11}=-i \tau/2$, $[\Sigma_{R}(\omega)]_{NN}=-i \tau/2$, and $[\Sigma_{P_n}(\omega)]_{nn}=-i \tau_{P}\nu^{(n)}/2$.  We will consider both the case of constant coupling to BVP, i.e, $\nu^{(n)}=1$, and the case of disordered coupling to BVP, where $\nu^{(n)}>0$, but otherwise random. Let us also define the average coupling to the probes $\overline{\tau}_P=\tau_P \overline{\nu}$, where $\overline{\nu}$ is the average value of $\nu^{(n)}$. 

To describe, conductance, we choose, $\mu_R=\epsilon_F$, $\mu_L=\epsilon_F+\delta \mu$, $\mu_n=\epsilon_F+\delta \mu_n$, where $\epsilon_F$ is the Fermi energy. The  conductance is then given by Landauer-B\"uttiker formula as \cite{buttiker_paper,molecular_buttiker,molecular_dephasing1, molecular_dephasing2,  molecular_dephasing4,segal-probe-1,segal-probe-2,segal-probe-3,malay,Bedkihal2013,Pastawski,SM}
\begin{align}
\label{conductance}
&\mathcal{G}(\epsilon_F)  = \tau^2  \left| G_{1N}(\epsilon_F) \right|^2 \nonumber \\
& +   \tau^2\tau_P \sum_{n,j=1}^{N}\nu^{(j)}\left|G_{Nn} (\epsilon_F)\right|^2 
\mathcal{W}^{-1}_{nj}(\epsilon_F) \left|G_{j1} (\epsilon_F)\right|^2 \hspace*{-2pt}. 
\end{align}
Here the elements of the  $N\times N$ matrix $\mathcal{W}(\epsilon_F)$ are
\begin{align}
\label{W}
\mathcal{W}_{nj} &= -\tau_P \nu^{(j)} |G_{nj}|^2, \quad \forall \, n\neq j \nonumber \\
\mathcal{W}_{nn} & =\tau (|G_{n1}|^2 +|G_{nN}|^2)+\tau_P  \sum_{j\neq n}^{N}\nu^{(j)}|G_{nj}|^2,
\end{align}
where we have suppressed the argument $\epsilon_F$  for brevity. The above equations show that, knowing the retarded NEGF, the conductance in presence of the probes can be obtained. In absence of the probes [i.e., setting $\tau_P=0$], the conductance is given by $\mathcal{G}^0(\epsilon_F)=\tau^2 \left| G_{1N}^0(\epsilon_F) \right|^2$, where $G^0(\epsilon_F)$ is the retarded NEGF in absence of the probes.

{\it Main result ---}
Let $\omega_b^{\pm}$ be the band edges of the system in the thermodynamic limit in absence of disorder. For our system, $\omega_b^{\pm}=\pm 2g$.  Then,  our main result can be succinctly stated as follows,
\begin{align}
\label{main_result}
& \mathcal{G}(\omega_b^{\pm}\mp \eta)~\textrm{increases monotonically with $N$} \nonumber \\
& \forall~~N_{\rm SB}^{(1)}< N < \min\{N_\eta,N_\varepsilon,N_{\beta},N_{\rm SB}^{(2)}\}.
\end{align}
Here $N_{\rm SB}^{(1)}\sim \overline{\tau}_P^{-1/3}$, $N_{\rm SB}^{(2)} \sim \overline{\tau}_P ^{-1/2}$ depend only on the average coupling to the probes, $N_\eta \sim \pi \eta^{-1/2}$ depends only on deviation from band edge, $N_\varepsilon \sim \sqrt{3} \pi \varepsilon^{-1/2}$ depends only on strength of disorder in the system, $N_{\beta} \sim \pi \beta^{1/2}$ depends only on temperature. Knowing these dependencies on various parameters, it is clear that this regime can be parametrically expanded by reducing $\overline{\tau}_P$, $\varepsilon$, $\eta$ and temperature. Thus, over a finite but extended regime of system lengths, there can be a superballistic scaling of conductance. Note that, the possibility of such behavior, even in an idealized setting, was not known before. In the following, we show that this fascinating behavior is a consequence of EP of transfer matrix occurring at every band edge. We consider the upper band edge ($\epsilon_F=\omega_b^{+}$). Due to particle-hole symmetry of the set-up, exactly same results are obtained at the lower band edge ($\epsilon_F=\omega_b^{-}$).
Henceforth, we set the system hopping parameter to $g=1$, which therefore sets our energy scale.

\begin{figure}
\includegraphics[width=\columnwidth]{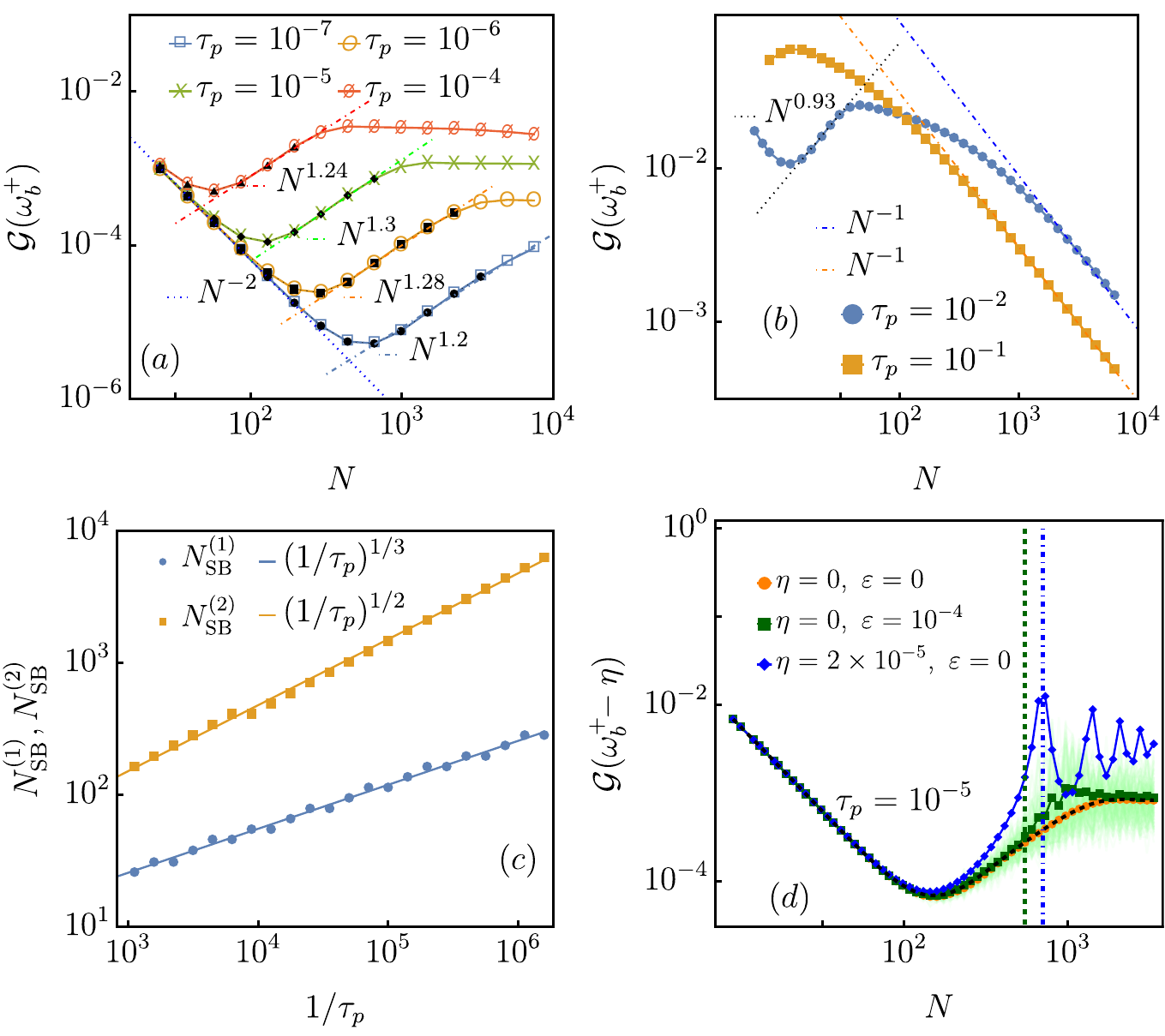} 
\caption{ (a) Behavior of $\mathcal{G}(\omega_b^{+})$ with system length $N$ for small values of $\tau_P$, without any disorder. The black dots show approximate result obtained on replacing $G(\omega_b^{+})$ by $G^0(\omega_b^+)$ (i.e., without the probes) in Eqs.~\eqref{conductance} and \eqref{W}. (b) Similar plots but at larger values of $\tau_P$ which captures the eventual crossover to conventional diffusive regime at large $N$. (c) The scaling of the start (end) of the superballistic regime $N_{\rm SB}^{(1)}$ ($N_{\rm SB}^{(2)}$) with $\tau_P^{-1}$, the continuous line showing fit of $\tau_P^{-1/3}$ ($\tau_P^{-1/2}$).   (d) The orange circles show $\mathcal{G}(\omega_b^{+})$ with disordered but weak coupling to BVPs ($\tau_P=10^{-5}$, $\nu^{(n)}$ randomly chosen between $0$ and $1$). The green squares show the same, with additional small disorder in the system ($\varepsilon=10^{-4}$). The light green continuous lines show results for individual disorder realizations. The blue diamonds show conductance versus $N$, without any disorder in the system, but with slight shift in Fermi energy from band edge ($\eta=2\times 10^{-5}$). The black dotted plot shows  $\mathcal{G}(\omega_b^{+})$ without any disorder, but with $\tau_P \to \overline{\tau}_P=0.5\tau_P$. The vertical green dashed line corresponds to $\sqrt{3}\pi \varepsilon^{-1/2}$ with $\varepsilon=10^{-4}$. The vertical blue dash-dotted line corresponds to $\pi \eta^{-1/2}$, with $\eta=2\times 10^{-5}$. 
} 
\label{fig1} 
\end{figure}

{\it Without disorder, exactly at band edge, zero temperature: numerical results---} First, we present numerical results in the complete absence of disorder, i.e., $\varepsilon=0$, constant coupling to BVP, $\nu^{(n)}=1$,  take $\beta \to \infty$ and Fermi energy exactly at upper band edge, $\epsilon_F=\omega_b^{+}$.   In Fig.~\ref{fig1}(a), we show plots of conductance with system length, for various small values of probe strength $\tau_P$. For small $N$, we clearly see a remnant of the subdiffusive scaling, $\mathcal{G}(\omega_b^{+}) \sim N^{-2}$ expected in absence of probes \cite{saha2022universal}. After this, we find the surprising superballistic regime, $\mathcal{G}(\omega_b^{+}) \sim N^\phi$, $\phi>0$ for a finite regime in system length. The exponent $\phi$ is non-universal. Importantly, the superballistic regime expands as $\tau_P$ is reduced. Beyond the superballistic regime, the conductance starts saturating with system length, eventually decaying as we increase the system length further. Although not seen in our numerics for small $\tau_P$ up to the largest accessible $N$, we expect this slow decay with system length to eventually lead to standard diffusive behavior $\mathcal{G}(\omega_b^{+})\sim N^{-1}$. This is captured for sufficiently large values of $\tau_P$ in Fig.~\ref{fig1}(b). 

To further analyze the superballistic regime, we extract from our numerics the onset and the termination of this regime. These correspond to the minimum and the following maximum of the plots in Fig.~\ref{fig1}(a), respectively. In Fig.~\ref{fig1}(c), we plot the starting (ending) system size of superballistic regime, $N_{\rm SB}^{(1)}$ ($N_{\rm SB}^{(2)}$), as a function of $\tau_P^{-1}$. We find that $N_{\rm SB}^{(1)}\sim \tau_P^{-1/3}$ and $N_{\rm SB}^{(2)}\sim \tau_P^{-1/2}$. For $\tau_P \ll 1$, we have  $\tau_P^{-1/3} \ll \tau_P^{-1/2}$, which shows that the superballistic regime can be enhanced by reducing $\tau_P$. Therefore, the extent of this superballistic regime, $N_{\rm SB}=N_{\rm SB}^{(2)}-N_{\rm SB}^{(1)}$ increases as $N_{\rm SB}\sim \tau_P^{-1/2}$, with decrease in $\tau_P$. 

{\it Origin of superballistic scaling: transfer matrix EPs ---} 
 For nearest neighbour one-dimensional systems, the retarded Green's function $G(\omega)$ is the inverse of a tridiagonal matrix. Using properties of tridiagonal matrices, in absence of any disorder, the elements of  $G(\omega)$ can be written as \cite{SM}
$
G_{\ell j}(\omega)=(-1)^{\ell+j} \frac{\Delta_{1,\ell-1}(\omega) \Delta_{N-j,N}(\omega)}{\Delta_{1,N}(\omega)},
$
where $\Delta_{1,\ell-1}(\omega), \Delta_{N-j,N}(\omega), \Delta_{1,N}(\omega)$ satisfy the following equations 
\begin{eqnarray}
\begin{pmatrix}
\Delta_{1,N}(\omega)  \\
\Delta_{1,N-1}(\omega)
\end{pmatrix} 
&=& 
\begin{pmatrix}
1 & \frac{i\tau}{2}  \\
0 & 1
\end{pmatrix} 
[\mathbf{T}(\omega)]^N
\begin{pmatrix}
 1 \\
 -\frac{i\tau}{2}
\end{pmatrix}, \nonumber \\
\begin{pmatrix}
\Delta_{1,\ell-1}(\omega)  \\
\Delta_{1,\ell-2} (\omega)
\end{pmatrix} &=& \begin{pmatrix}
1 & \frac{i\tau}{2}  \\
0 & 1
\end{pmatrix} 
[\mathbf{T}(\omega)]^{\ell-1}  
\begin{pmatrix}
 1 \\
 0
\end{pmatrix},\\
\begin{pmatrix}
\Delta_{N-j,N}(\omega) \\
\Delta_{N-j-1,N}(\omega)
\end{pmatrix} &=&  
[\mathbf{T}(\omega)]^{N-j}  
\begin{pmatrix}
 1 \\
 -\frac{i\tau}{2}
\end{pmatrix}. \nonumber 
\label{determinants}
\end{eqnarray}
In above $\mathbf{T}(\omega)$ is a $2\times2$ matrix given by
$
\mathbf{T}(\omega)= \mathbf{T}^0(\omega) +  \frac{i\tau_P}{4}\left(\mathbb{I}_2+\sigma_z\right),  
$
where $\mathbf{T}^0(\omega)$=$\frac{\omega}{2}\left(\mathbb{I}_2+\sigma_z\right)-i\sigma_y$ is the transfer matrix of the tight-binding chain, $\mathbb{I}_2$ is $2\times2$ identity matrix and $\sigma_{x,y,z}$ are the Pauli matrices. 
The above equations show that the nature of $\mathbf{T}(\omega)$ controls the system size scaling of various elements of the retarded NEGF in presence of BVPs, while $\mathbf{T}^{0}(\omega)$ does the same in their absence. It can be easily checked that  $\mathbf{T}^{0}(\omega)$ has EPs at $\omega=\omega_b^{\pm}$. As has been recently shown \cite{saha2022universal}, this behavior is a consequence of an antilinear symmetry of transfer matrices of nearest-neighbour tight-binding chains, which makes them pseudo-Hermitian. 
It holds in general even for multi-band cases. As a consequence, in absence of the probes, conductance shows a universal subdiffusive scaling $\mathcal{G}^0(\omega_b^{+})\sim N^{-2}$ at every band edge \cite{saha2022universal}. In our plots in Fig.~\ref{fig1}, this is seen to survive up to a finite size, $N_{\rm SB}^{(1)}$,  for small $\tau_P$.

For small $\tau_P$ and $N\ll N_{\rm SB}^{(2)}$ the leading order behavior should be captured by using $\mathbf{T}(\omega_b^{+}) \simeq \mathbf{T}^0(\omega_b^{+})$. This is same as using Eqs.\eqref{conductance}, \eqref{W}, with $G(\epsilon_F)$ replaced by $G^0(\epsilon_F)$. This approximation can also be justified using a more careful, order-by-order perturbation in $\tau_P$ \cite{SM}. Conductance calculated in this approximation is shown by the black dots in Fig.~\ref{fig1}(a). Indeed they overlap with the exact results in the entire subdiffusive and the superballistic regimes. This clearly establishes that the superballistic regimes stems from how the second term in Eq.\eqref{conductance}, which embodies the effect of BVPs, is affected by the transfer matrix EP occurring at the band edge. 

To explain the scaling of $N_{\rm SB}^{(1)}$ with $\tau_P$, we look at the condition for observing the subdiffusive scaling. Clearly, this is seen in the regime where the effect of the probes is negligible. So, we calculate the expression for conductance up to the lowest order in $\tau_P$. Since the second term in Eq.~\eqref{conductance} is explicitly proportional to $\tau_P$, in calculating all required matrix elements for that term, we simply set $\tau_P$ to zero. From Eq.~\eqref{W}, we see that this makes $\mathcal{W}(\epsilon_F)$ diagonal, which leads to 
$
\mathcal{G}= \tau^2  \left| G_{1N}^0 \right|^2+\tau \tau_P \sum_{\ell=1}^{N} \! \frac{|G_{\ell N}^0|^2  |G_{1\ell}^0|^2}{|G_{1\ell}^0|^2+|G_{\ell N}^0|^2} + \mathcal{O}(\tau_P^2), 
$
where we have suppressed the argument $\epsilon_F$ for brevity. Due to transfer matrix EP at $\epsilon_F=\omega_b^{+}$, it can be checked that the second term in above expression diverges as $N$, while the first term decays as $N^{-2}$ \cite{SM}. Clearly, $N_{\rm SB}^{(1)}$, which gives the end of the subdiffusive regime and the beginning of the superballistic regime, must correspond to the case where the two terms are comparable,   
$\tau{N_{\rm SB}^{(1)}}^{-2} \sim \tau_P N_{\rm SB}^{(1)}$. This directly gives $N_{\rm SB}^{(1)} \sim \tau_P^{-1/3}$, as has been numerically seen in Fig.~\ref{fig1}(c). Thus, we conclude that the superballistic scaling comes from the second term in Eq.~\eqref{conductance}, in the regime where the EP of the transfer matrix governs its leading behavior.  

To explain the scaling of $N_{\rm SB}^{(2)}$ with $\tau_P$, we note that the presence of the probes makes $\mathbf{T}(\omega)$ always diagonalizable, with magnitude of one of the eigenvalues $>1$, as can be easily confirmed by direct calculation. Consequently, it can be shown that, for $|\ell - j|$ large, $|G_{\ell j}(\omega)|^2\sim e^{-|\ell-j|/\xi}$, with $\xi^{-1}=2\kappa_1=2\log |\lambda_+|$ \cite{SM} where $\lambda_+$ is the eigenvalue of $\mathbf{T}(\omega)$ with higher magnitude. Whenever this is the case, the conductance in Eq~.\eqref{conductance} gives diffusive scaling $\mathcal{G}(\omega)\sim N^{-1}$, for $N\gg \xi$, as has been shown in seminal work \cite{Pastawski}.
This holds for all values of $\omega$. Hence, in presence of probes, at large enough system sizes, we always get diffusive behavior, as shown in Fig.~\ref{fig1}(b). Thus, the superballistic scaling is observed for $N\lesssim \xi$. So, we have $N_{\rm SB}^{(2)}\sim \xi$.
At band edges, to leading order on $\tau_P$, the magnitude of eigenvalues of $\mathbf{T}(\omega)$ can be shown to be $|\lambda_{\pm}|\simeq 1\pm \sqrt{\tau_P}$. This leads to $\xi^{-1}=2\log|\lambda_+|\simeq 2\sqrt{\tau_P}$ \cite{SM}, which then gives $N_{\rm SB}^{(2)} \sim \tau_P^{-1/2}$, as numerically seen in Fig.~\ref{fig1}(c).

{\it Effect of small disorder, small shifts from band edge, finite temperature ---}
If disordered couplings to BVPs are considered, we numerically see that up to the superballistic regime, the conductance is almost exactly same as the case of uniform coupling with the average strength, i.e, $\tau_P \to \overline{\tau}_P$. Figure~\ref{fig1}(d) (orange circles) shows a representative plot, which completely overlaps with the corresponding uniform coupling result (black dotted line).

Disorder in the system on-site energies and hoppings, i.e, when $\varepsilon> 0$, non-perturbatively affects the single particle eigenstates and induces localization. But, for weak disorder, up to a  finite system-size, the eigenstates are affected only perturbatively, and the effects of localization do not manifest. It is within this regime that we expect the conductance scaling to also remain almost unaffected. Since, we are considering the upper band edge, we look for the system size up to which perturbation theory holds for highest single particle eigenstate.  For the tight-binding chain, this system size can be calculated as $N_\varepsilon \sim \sqrt{3} \pi \varepsilon^{-1/2}$. Below $N_\varepsilon$ we expect negligible effect of disorder. Figure~\ref{fig1}(d) (compare green squares with black dotted line) shows a representative plot with numerical evidence of this.

The transfer matrix EPs are at the band edges of the system in the thermodynamic limit. At any finite system size $N$, let the maximum single-particle eigenvalue be $\omega_N^{+}$. We have $\omega_b^+-\omega_N^{+}\simeq \pi^2 N^{-2}$ for the tight-binding chain. In absence of any BVPs, it can be checked that the subdiffusive scaling of conductance holds when $\omega_N^{+} < \epsilon_F \leq \omega_b^{+}$. Since, the superballistic regime occurs due to effect of the BVPs on the subdiffusive scaling regime, beyond this regime the superballistic scaling cannot be expected. For $\epsilon_F=\omega_b^{+}-\eta$, with given $\eta$, this gives the length scale depending only of $\eta$, $N_\eta\sim \pi \eta^{-1/2}$, beyond which conductance should not be monotonically increasing. This behavior is shown in Fig.~\ref{fig1}(d) (blue diamonds). We see that superballistic scaling function is drastically changed on shifting $\epsilon_F$. However, note that, even at  $\epsilon_F=\omega_b^+$,  the scaling function was not universal (see Fig.~\ref{fig1}(a)).

Small finite temperature changes the expression for conductance \cite{buttiker_paper,molecular_buttiker,molecular_dephasing1, molecular_dephasing2,  molecular_dephasing4,segal-probe-1,segal-probe-2,segal-probe-3,malay,Bedkihal2013,Pastawski}, leading to an integration over energies around $\epsilon_F$ in a width $\sim 1/\beta$. So, by similar arguments as above,  superballistic regime is expected to hold if temperature satisfies $\beta (\omega_b^+-\omega_N^{+})\simeq \beta\pi^2 N^{-2} \gg 1$. For a chosen $\beta$, this gives the system-size $N_{\beta}\sim \pi \beta^{1/2}$ beyond which the superballistic regime is not expected. However, due to numerical instabilities, this regime is difficult to access computationally.

Combining all of the above, we arrive at our main result in Eq.\eqref{main_result}. We reiterate that there is no assumption on strength of coupling to the source-drain leads, $\tau$, which need not be small. Below, we further generalize the result.

{\it Generalization to multi-band systems ---}
Our analytical understanding shows that the superballistic regime stems from an interplay of the EPs of the transfer matrix, and the presence of BVPs. There is a transfer matrix EP at every band edge of any finite-ranged tight-binding chain, whether single-band or multi-band \cite{saha2022universal}. Consequently, a superballistic regime will be seen near every band edge in all such cases. The multi-band case can arise from presence of an additional periodic on-site potential in the system. Taking such a two-band case, the superballistic scaling can be easily explicitly confirmed \cite{SM}.

{\it Probes versus many-body interaction ---} 
Akin to BVPs, many-body interactions in the system can lead to inelastic scattering processes \cite{molecular_buttiker, buttikerdephase, electron-phonon,electron_phonon2}. But, the superballistic scaling cannot be obtained with many-body interactions alone. This is because, at band edges, it can be argued that number conserving many-body interactions have negligible effect due to vanishing particle or hole density \cite{SM}. Therefore, presence of surrounding environment is crucial for superballistic scaling. The simultaneous presence of BVPs and many-body interactions remains a challenging and interesting question beyond present analytical and numerical techniques.

{\it Conclusions and outlook ---}
We reveal how non-Hermitian properties of the transfer matrix affect environment-assisted transport, leading to superballistic scaling of conductance, a completely different regime of quantum transport. Physical effects of EPs of non-Hermitian matrices are of interest in the field of non-Hermitian physics and optics \cite{Feng2017,El_Ganainy2018,El_Ganainy2019,Ashida_2020}. Different kinds of anomalous transport and their microscopic origins are of interest in statistical physics \cite{Landi_2022,Bertini_2021,Dhar_2019,Xu_2016, Dhar_2008}. Environment-assisted transport is of interest in fields of mesoscopic physics \cite{buttikerdephase,segal-probe-1,segal-probe-2,segal-probe-3,dephasing_dubi,Landi-latest,ENQT1,diffusion1,buttikerdephase, saha_buttiker}, quantum many-body physics \cite{dephasing-interaction1, dephasing_spin_interaction2, dephasing_interaction3, Mendoza-Arenas_2013}, quantum chemistry and biology \cite{Plenio_2008,Rebentrost_2009,Caruso_2009, Sowa_2017,Harush_2018,Dutta_2017}, quantum thermodynamics \cite{buttikerdephase,saha_buttiker} and also in quantum simulation experiments \cite{Maier_2019,Leon-Montiel_2015,Biggerstaff_2016,Viciani_2015,Harris_2017}. Our results therefore connect these different research directions. The practical implications of our results can be investigated in nanoscale quantum systems treated within density functional theory \cite{molecular_buttiker,Brandbyge_2002,Nitzan_2003,Hjorth_2017,Kurth_2019,Smidstrup_2020}, as well as in two-dimensional topological insulators \cite{Kou_2017, Hasan_2010} which provide effective one-dimensional ballistic transport channels.

\begin{acknowledgements}
M. S. acknowledges financial support through National Postdoctoral Fellowship (NPDF), SERB file no.~PDF/2020/000992. B. K. A. acknowledges the MATRICS grant (MTR/2020/000472) from SERB, Government of India, and the Shastri Indo-Canadian Institute for providing financial support for this research work in the form of a Shastri Institutional Collaborative Research Grant (SICRG). 
M.K. would like to acknowledge support from the project 6004-1 of the Indo-French Centre for the Promotion of Advanced Research (IFCPAR), Ramanujan Fellowship (SB/S2/RJN-114/2016), SERB Early Career Research Award (ECR/2018/002085) and SERB Matrics Grant (MTR/2019/001101) from the Science and Engineering Research Board (SERB), Department of Science and Technology, Government of India. MK acknowledges support of the Department of Atomic Energy, Government of India, under Project No. RTI4001. A.P acknowledges funding from the European Union’s Horizon 2020 research and innovation programme under the Marie Sklodowska-Curie Grant Agreement No. 890884. A.P also acknowledges funding from the Danish National Research Foundation through the Center of Excellence ``CCQ'' (Grant agreement no.: DNRF156). This research was supported in part by the International Centre for Theoretical Sciences (ICTS) for participating in the program - Physics with Trapped Atoms, Molecules and Ions (code: ICTS/TAMIONs-2022/5). M.K. acknowledges support from the Infosys Foundation International Exchange Program at ICTS.
\end{acknowledgements}

\bibliography{bandedge_probe}

\begin{thebibliography}{78}%
\makeatletter
\providecommand \@ifxundefined [1]{%
 \@ifx{#1\undefined}
}%
\providecommand \@ifnum [1]{%
 \ifnum #1\expandafter \@firstoftwo
 \else \expandafter \@secondoftwo
 \fi
}%
\providecommand \@ifx [1]{%
 \ifx #1\expandafter \@firstoftwo
 \else \expandafter \@secondoftwo
 \fi
}%
\providecommand \natexlab [1]{#1}%
\providecommand \enquote  [1]{``#1''}%
\providecommand \bibnamefont  [1]{#1}%
\providecommand \bibfnamefont [1]{#1}%
\providecommand \citenamefont [1]{#1}%
\providecommand \href@noop [0]{\@secondoftwo}%
\providecommand \href [0]{\begingroup \@sanitize@url \@href}%
\providecommand \@href[1]{\@@startlink{#1}\@@href}%
\providecommand \@@href[1]{\endgroup#1\@@endlink}%
\providecommand \@sanitize@url [0]{\catcode `\\12\catcode `\$12\catcode
  `\&12\catcode `\#12\catcode `\^12\catcode `\_12\catcode `\%12\relax}%
\providecommand \@@startlink[1]{}%
\providecommand \@@endlink[0]{}%
\providecommand \url  [0]{\begingroup\@sanitize@url \@url }%
\providecommand \@url [1]{\endgroup\@href {#1}{\urlprefix }}%
\providecommand \urlprefix  [0]{URL }%
\providecommand \Eprint [0]{\href }%
\providecommand \doibase [0]{http://dx.doi.org/}%
\providecommand \selectlanguage [0]{\@gobble}%
\providecommand \bibinfo  [0]{\@secondoftwo}%
\providecommand \bibfield  [0]{\@secondoftwo}%
\providecommand \translation [1]{[#1]}%
\providecommand \BibitemOpen [0]{}%
\providecommand \bibitemStop [0]{}%
\providecommand \bibitemNoStop [0]{.\EOS\space}%
\providecommand \EOS [0]{\spacefactor3000\relax}%
\providecommand \BibitemShut  [1]{\csname bibitem#1\endcsname}%
\let\auto@bib@innerbib\@empty
\bibitem [{\citenamefont {Xu}\ \emph {et~al.}(2014)\citenamefont {Xu},
  \citenamefont {Pereira}, \citenamefont {Wang}, \citenamefont {Wu},
  \citenamefont {Zhang}, \citenamefont {Zhao}, \citenamefont {Bae},
  \citenamefont {Tinh~Bui}, \citenamefont {Xie}, \citenamefont {Thong},
  \citenamefont {Hong}, \citenamefont {Loh}, \citenamefont {Donadio},
  \citenamefont {Li},\ and\ \citenamefont {{\"O}zyilmaz}}]{length_dependent}%
  \BibitemOpen
  \bibfield  {author} {\bibinfo {author} {\bibfnamefont {X.}~\bibnamefont
  {Xu}}, \bibinfo {author} {\bibfnamefont {L.~F.~C.}\ \bibnamefont {Pereira}},
  \bibinfo {author} {\bibfnamefont {Y.}~\bibnamefont {Wang}}, \bibinfo {author}
  {\bibfnamefont {J.}~\bibnamefont {Wu}}, \bibinfo {author} {\bibfnamefont
  {K.}~\bibnamefont {Zhang}}, \bibinfo {author} {\bibfnamefont
  {X.}~\bibnamefont {Zhao}}, \bibinfo {author} {\bibfnamefont {S.}~\bibnamefont
  {Bae}}, \bibinfo {author} {\bibfnamefont {C.}~\bibnamefont {Tinh~Bui}},
  \bibinfo {author} {\bibfnamefont {R.}~\bibnamefont {Xie}}, \bibinfo {author}
  {\bibfnamefont {J.~T.~L.}\ \bibnamefont {Thong}}, \bibinfo {author}
  {\bibfnamefont {B.~H.}\ \bibnamefont {Hong}}, \bibinfo {author}
  {\bibfnamefont {K.~P.}\ \bibnamefont {Loh}}, \bibinfo {author} {\bibfnamefont
  {D.}~\bibnamefont {Donadio}}, \bibinfo {author} {\bibfnamefont
  {B.}~\bibnamefont {Li}}, \ and\ \bibinfo {author} {\bibfnamefont
  {B.}~\bibnamefont {{\"O}zyilmaz}},\ }\href {\doibase 10.1038/ncomms4689}
  {\bibfield  {journal} {\bibinfo  {journal} {Nature Communications}\ }\textbf
  {\bibinfo {volume} {5}},\ \bibinfo {pages} {3689} (\bibinfo {year}
  {2014})}\BibitemShut {NoStop}%
\bibitem [{\citenamefont {Meier}\ \emph {et~al.}(2014)\citenamefont {Meier},
  \citenamefont {Menges}, \citenamefont {Nirmalraj}, \citenamefont
  {H\"olscher}, \citenamefont {Riel},\ and\ \citenamefont
  {Gotsmann}}]{length_thermal_molecular_chain}%
  \BibitemOpen
  \bibfield  {author} {\bibinfo {author} {\bibfnamefont {T.}~\bibnamefont
  {Meier}}, \bibinfo {author} {\bibfnamefont {F.}~\bibnamefont {Menges}},
  \bibinfo {author} {\bibfnamefont {P.}~\bibnamefont {Nirmalraj}}, \bibinfo
  {author} {\bibfnamefont {H.}~\bibnamefont {H\"olscher}}, \bibinfo {author}
  {\bibfnamefont {H.}~\bibnamefont {Riel}}, \ and\ \bibinfo {author}
  {\bibfnamefont {B.}~\bibnamefont {Gotsmann}},\ }\href {\doibase
  10.1103/PhysRevLett.113.060801} {\bibfield  {journal} {\bibinfo  {journal}
  {Phys. Rev. Lett.}\ }\textbf {\bibinfo {volume} {113}},\ \bibinfo {pages}
  {060801} (\bibinfo {year} {2014})}\BibitemShut {NoStop}%
\bibitem [{\citenamefont {Majee}\ and\ \citenamefont
  {Aksamija}(2016)}]{divergent_thermal_conductivity_graphene}%
  \BibitemOpen
  \bibfield  {author} {\bibinfo {author} {\bibfnamefont {A.~K.}\ \bibnamefont
  {Majee}}\ and\ \bibinfo {author} {\bibfnamefont {Z.}~\bibnamefont
  {Aksamija}},\ }\href {\doibase 10.1103/PhysRevB.93.235423} {\bibfield
  {journal} {\bibinfo  {journal} {Phys. Rev. B}\ }\textbf {\bibinfo {volume}
  {93}},\ \bibinfo {pages} {235423} (\bibinfo {year} {2016})}\BibitemShut
  {NoStop}%
\bibitem [{\citenamefont {Li}\ \emph {et~al.}(2017)\citenamefont {Li},
  \citenamefont {Takahashi},\ and\ \citenamefont
  {Zhang}}]{divergent_thermal_conductivity}%
  \BibitemOpen
  \bibfield  {author} {\bibinfo {author} {\bibfnamefont {Q.-Y.}\ \bibnamefont
  {Li}}, \bibinfo {author} {\bibfnamefont {K.}~\bibnamefont {Takahashi}}, \
  and\ \bibinfo {author} {\bibfnamefont {X.}~\bibnamefont {Zhang}},\ }\href
  {\doibase 10.1103/PhysRevLett.119.179601} {\bibfield  {journal} {\bibinfo
  {journal} {Phys. Rev. Lett.}\ }\textbf {\bibinfo {volume} {119}},\ \bibinfo
  {pages} {179601} (\bibinfo {year} {2017})}\BibitemShut {NoStop}%
\bibitem [{\citenamefont {Nika}\ \emph {et~al.}(2012)\citenamefont {Nika},
  \citenamefont {Askerov},\ and\ \citenamefont
  {Balandin}}]{size_thermal_graphene}%
  \BibitemOpen
  \bibfield  {author} {\bibinfo {author} {\bibfnamefont {D.~L.}\ \bibnamefont
  {Nika}}, \bibinfo {author} {\bibfnamefont {A.~S.}\ \bibnamefont {Askerov}}, \
  and\ \bibinfo {author} {\bibfnamefont {A.~A.}\ \bibnamefont {Balandin}},\
  }\href {\doibase 10.1021/nl301230g} {\bibfield  {journal} {\bibinfo
  {journal} {Nano Letters}\ }\textbf {\bibinfo {volume} {12}},\ \bibinfo
  {pages} {3238} (\bibinfo {year} {2012})}\BibitemShut {NoStop}%
\bibitem [{\citenamefont {Purkayastha}\ \emph {et~al.}(2017)\citenamefont
  {Purkayastha}, \citenamefont {Dhar},\ and\ \citenamefont
  {Kulkarni}}]{Manas-anomalous-1}%
  \BibitemOpen
  \bibfield  {author} {\bibinfo {author} {\bibfnamefont {A.}~\bibnamefont
  {Purkayastha}}, \bibinfo {author} {\bibfnamefont {A.}~\bibnamefont {Dhar}}, \
  and\ \bibinfo {author} {\bibfnamefont {M.}~\bibnamefont {Kulkarni}},\ }\href
  {\doibase 10.1103/PhysRevB.96.180204} {\bibfield  {journal} {\bibinfo
  {journal} {Phys. Rev. B}\ }\textbf {\bibinfo {volume} {96}},\ \bibinfo
  {pages} {180204} (\bibinfo {year} {2017})}\BibitemShut {NoStop}%
\bibitem [{\citenamefont {Purkayastha}\ \emph {et~al.}(2018)\citenamefont
  {Purkayastha}, \citenamefont {Sanyal}, \citenamefont {Dhar},\ and\
  \citenamefont {Kulkarni}}]{Manas-anomalous-2}%
  \BibitemOpen
  \bibfield  {author} {\bibinfo {author} {\bibfnamefont {A.}~\bibnamefont
  {Purkayastha}}, \bibinfo {author} {\bibfnamefont {S.}~\bibnamefont {Sanyal}},
  \bibinfo {author} {\bibfnamefont {A.}~\bibnamefont {Dhar}}, \ and\ \bibinfo
  {author} {\bibfnamefont {M.}~\bibnamefont {Kulkarni}},\ }\href {\doibase
  10.1103/PhysRevB.97.174206} {\bibfield  {journal} {\bibinfo  {journal} {Phys.
  Rev. B}\ }\textbf {\bibinfo {volume} {97}},\ \bibinfo {pages} {174206}
  (\bibinfo {year} {2018})}\BibitemShut {NoStop}%
\bibitem [{\citenamefont {Xu}(2016)}]{Xu_2016}%
  \BibitemOpen
  \bibfield  {author} {\bibinfo {author} {\bibfnamefont {Z.}~\bibnamefont
  {Xu}},\ }\href {\doibase https://doi.org/10.1016/j.taml.2016.04.002}
  {\bibfield  {journal} {\bibinfo  {journal} {Theoretical and Applied Mechanics
  Letters}\ }\textbf {\bibinfo {volume} {6}},\ \bibinfo {pages} {113} (\bibinfo
  {year} {2016})}\BibitemShut {NoStop}%
\bibitem [{\citenamefont {Dhar}\ \emph {et~al.}(2019)\citenamefont {Dhar},
  \citenamefont {Kundu},\ and\ \citenamefont {Kundu}}]{Dhar_2019}%
  \BibitemOpen
  \bibfield  {author} {\bibinfo {author} {\bibfnamefont {A.}~\bibnamefont
  {Dhar}}, \bibinfo {author} {\bibfnamefont {A.}~\bibnamefont {Kundu}}, \ and\
  \bibinfo {author} {\bibfnamefont {A.}~\bibnamefont {Kundu}},\ }\href
  {\doibase 10.3389/fphy.2019.00159} {\bibfield  {journal} {\bibinfo  {journal}
  {Frontiers in Physics}\ }\textbf {\bibinfo {volume} {7}} (\bibinfo {year}
  {2019}),\ 10.3389/fphy.2019.00159}\BibitemShut {NoStop}%
\bibitem [{\citenamefont {Dhar}(2008)}]{Dhar_2008}%
  \BibitemOpen
  \bibfield  {author} {\bibinfo {author} {\bibfnamefont {A.}~\bibnamefont
  {Dhar}},\ }\href@noop {} {\bibfield  {journal} {\bibinfo  {journal} {Advances
  in Physics}\ }\textbf {\bibinfo {volume} {57}},\ \bibinfo {pages} {457}
  (\bibinfo {year} {2008})}\BibitemShut {NoStop}%
\bibitem [{\citenamefont {Bertini}\ \emph {et~al.}(2021)\citenamefont
  {Bertini}, \citenamefont {Heidrich-Meisner}, \citenamefont {Karrasch},
  \citenamefont {Prosen}, \citenamefont {Steinigeweg},\ and\ \citenamefont
  {\ifmmode \check{Z}\else \v{Z}\fi{}nidari\ifmmode~\check{c}\else
  \v{c}\fi{}}}]{Bertini_2021}%
  \BibitemOpen
  \bibfield  {author} {\bibinfo {author} {\bibfnamefont {B.}~\bibnamefont
  {Bertini}}, \bibinfo {author} {\bibfnamefont {F.}~\bibnamefont
  {Heidrich-Meisner}}, \bibinfo {author} {\bibfnamefont {C.}~\bibnamefont
  {Karrasch}}, \bibinfo {author} {\bibfnamefont {T.}~\bibnamefont {Prosen}},
  \bibinfo {author} {\bibfnamefont {R.}~\bibnamefont {Steinigeweg}}, \ and\
  \bibinfo {author} {\bibfnamefont {M.}~\bibnamefont {\ifmmode \check{Z}\else
  \v{Z}\fi{}nidari\ifmmode~\check{c}\else \v{c}\fi{}}},\ }\href {\doibase
  10.1103/RevModPhys.93.025003} {\bibfield  {journal} {\bibinfo  {journal}
  {Rev. Mod. Phys.}\ }\textbf {\bibinfo {volume} {93}},\ \bibinfo {pages}
  {025003} (\bibinfo {year} {2021})}\BibitemShut {NoStop}%
\bibitem [{\citenamefont {Landi}\ \emph {et~al.}(2022)\citenamefont {Landi},
  \citenamefont {Poletti},\ and\ \citenamefont {Schaller}}]{Landi_2022}%
  \BibitemOpen
  \bibfield  {author} {\bibinfo {author} {\bibfnamefont {G.~T.}\ \bibnamefont
  {Landi}}, \bibinfo {author} {\bibfnamefont {D.}~\bibnamefont {Poletti}}, \
  and\ \bibinfo {author} {\bibfnamefont {G.}~\bibnamefont {Schaller}},\ }\href
  {\doibase 10.1103/RevModPhys.94.045006} {\bibfield  {journal} {\bibinfo
  {journal} {Rev. Mod. Phys.}\ }\textbf {\bibinfo {volume} {94}},\ \bibinfo
  {pages} {045006} (\bibinfo {year} {2022})}\BibitemShut {NoStop}%
\bibitem [{\citenamefont {Li}\ \emph {et~al.}(2020)\citenamefont {Li},
  \citenamefont {Amado}, \citenamefont {Hyart}, \citenamefont {Mazur},\ and\
  \citenamefont {Robinson}}]{Li_2020}%
  \BibitemOpen
  \bibfield  {author} {\bibinfo {author} {\bibfnamefont {Y.}~\bibnamefont
  {Li}}, \bibinfo {author} {\bibfnamefont {M.}~\bibnamefont {Amado}}, \bibinfo
  {author} {\bibfnamefont {T.}~\bibnamefont {Hyart}}, \bibinfo {author}
  {\bibfnamefont {G.~P.}\ \bibnamefont {Mazur}}, \ and\ \bibinfo {author}
  {\bibfnamefont {J.~W.~A.}\ \bibnamefont {Robinson}},\ }\href {\doibase
  10.1038/s42005-020-00495-y} {\bibfield  {journal} {\bibinfo  {journal}
  {Communications Physics}\ }\textbf {\bibinfo {volume} {3}},\ \bibinfo {pages}
  {224} (\bibinfo {year} {2020})}\BibitemShut {NoStop}%
\bibitem [{\citenamefont {Homoth}\ \emph {et~al.}(2009)\citenamefont {Homoth},
  \citenamefont {Wenderoth}, \citenamefont {Druga}, \citenamefont {Winking},
  \citenamefont {Ulbrich}, \citenamefont {Bobisch}, \citenamefont {Weyers},
  \citenamefont {Bannani}, \citenamefont {Zubkov}, \citenamefont {Bernhart},
  \citenamefont {Kaspers},\ and\ \citenamefont {Möller}}]{Homoth_2009}%
  \BibitemOpen
  \bibfield  {author} {\bibinfo {author} {\bibfnamefont {J.}~\bibnamefont
  {Homoth}}, \bibinfo {author} {\bibfnamefont {M.}~\bibnamefont {Wenderoth}},
  \bibinfo {author} {\bibfnamefont {T.}~\bibnamefont {Druga}}, \bibinfo
  {author} {\bibfnamefont {L.}~\bibnamefont {Winking}}, \bibinfo {author}
  {\bibfnamefont {R.~G.}\ \bibnamefont {Ulbrich}}, \bibinfo {author}
  {\bibfnamefont {C.~A.}\ \bibnamefont {Bobisch}}, \bibinfo {author}
  {\bibfnamefont {B.}~\bibnamefont {Weyers}}, \bibinfo {author} {\bibfnamefont
  {A.}~\bibnamefont {Bannani}}, \bibinfo {author} {\bibfnamefont
  {E.}~\bibnamefont {Zubkov}}, \bibinfo {author} {\bibfnamefont {A.~M.}\
  \bibnamefont {Bernhart}}, \bibinfo {author} {\bibfnamefont {M.~R.}\
  \bibnamefont {Kaspers}}, \ and\ \bibinfo {author} {\bibfnamefont
  {R.}~\bibnamefont {Möller}},\ }\href {\doibase 10.1021/nl803783g} {\bibfield
   {journal} {\bibinfo  {journal} {Nano Letters}\ }\textbf {\bibinfo {volume}
  {9}},\ \bibinfo {pages} {1588} (\bibinfo {year} {2009})},\ \bibinfo {note}
  {pMID: 19278211}\BibitemShut {NoStop}%
\bibitem [{\citenamefont {Liu}\ \emph {et~al.}(2005)\citenamefont {Liu},
  \citenamefont {Stock},\ and\ \citenamefont {Rudolph}}]{Liu_2005}%
  \BibitemOpen
  \bibfield  {author} {\bibinfo {author} {\bibfnamefont {X.}~\bibnamefont
  {Liu}}, \bibinfo {author} {\bibfnamefont {R.}~\bibnamefont {Stock}}, \ and\
  \bibinfo {author} {\bibfnamefont {W.}~\bibnamefont {Rudolph}},\ }\href
  {\doibase 10.1103/PhysRevB.72.195431} {\bibfield  {journal} {\bibinfo
  {journal} {Phys. Rev. B}\ }\textbf {\bibinfo {volume} {72}},\ \bibinfo
  {pages} {195431} (\bibinfo {year} {2005})}\BibitemShut {NoStop}%
\bibitem [{\citenamefont {Li}\ \emph {et~al.}(2005)\citenamefont {Li},
  \citenamefont {Lu}, \citenamefont {Li}, \citenamefont {Bai},\ and\
  \citenamefont {Gu}}]{Li_2005}%
  \BibitemOpen
  \bibfield  {author} {\bibinfo {author} {\bibfnamefont {H.~J.}\ \bibnamefont
  {Li}}, \bibinfo {author} {\bibfnamefont {W.~G.}\ \bibnamefont {Lu}}, \bibinfo
  {author} {\bibfnamefont {J.~J.}\ \bibnamefont {Li}}, \bibinfo {author}
  {\bibfnamefont {X.~D.}\ \bibnamefont {Bai}}, \ and\ \bibinfo {author}
  {\bibfnamefont {C.~Z.}\ \bibnamefont {Gu}},\ }\href {\doibase
  10.1103/PhysRevLett.95.086601} {\bibfield  {journal} {\bibinfo  {journal}
  {Phys. Rev. Lett.}\ }\textbf {\bibinfo {volume} {95}},\ \bibinfo {pages}
  {086601} (\bibinfo {year} {2005})}\BibitemShut {NoStop}%
\bibitem [{\citenamefont {Debray}\ \emph {et~al.}(2000)\citenamefont {Debray},
  \citenamefont {Raichev}, \citenamefont {Vasilopoulos}, \citenamefont
  {Rahman}, \citenamefont {Perrin},\ and\ \citenamefont
  {Mitchell}}]{Debray_2000}%
  \BibitemOpen
  \bibfield  {author} {\bibinfo {author} {\bibfnamefont {P.}~\bibnamefont
  {Debray}}, \bibinfo {author} {\bibfnamefont {O.~E.}\ \bibnamefont {Raichev}},
  \bibinfo {author} {\bibfnamefont {P.}~\bibnamefont {Vasilopoulos}}, \bibinfo
  {author} {\bibfnamefont {M.}~\bibnamefont {Rahman}}, \bibinfo {author}
  {\bibfnamefont {R.}~\bibnamefont {Perrin}}, \ and\ \bibinfo {author}
  {\bibfnamefont {W.~C.}\ \bibnamefont {Mitchell}},\ }\href {\doibase
  10.1103/PhysRevB.61.10950} {\bibfield  {journal} {\bibinfo  {journal} {Phys.
  Rev. B}\ }\textbf {\bibinfo {volume} {61}},\ \bibinfo {pages} {10950}
  (\bibinfo {year} {2000})}\BibitemShut {NoStop}%
\bibitem [{\citenamefont {Feng}\ \emph {et~al.}(2017)\citenamefont {Feng},
  \citenamefont {El-Ganainy},\ and\ \citenamefont {Ge}}]{Feng2017}%
  \BibitemOpen
  \bibfield  {author} {\bibinfo {author} {\bibfnamefont {L.}~\bibnamefont
  {Feng}}, \bibinfo {author} {\bibfnamefont {R.}~\bibnamefont {El-Ganainy}}, \
  and\ \bibinfo {author} {\bibfnamefont {L.}~\bibnamefont {Ge}},\ }\href
  {\doibase 10.1038/s41566-017-0031-1} {\bibfield  {journal} {\bibinfo
  {journal} {Nature Photonics}\ }\textbf {\bibinfo {volume} {11}},\ \bibinfo
  {pages} {752} (\bibinfo {year} {2017})}\BibitemShut {NoStop}%
\bibitem [{\citenamefont {El-Ganainy}\ \emph {et~al.}(2018)\citenamefont
  {El-Ganainy}, \citenamefont {Makris}, \citenamefont {Khajavikhan},
  \citenamefont {Musslimani}, \citenamefont {Rotter},\ and\ \citenamefont
  {Christodoulides}}]{El_Ganainy2018}%
  \BibitemOpen
  \bibfield  {author} {\bibinfo {author} {\bibfnamefont {R.}~\bibnamefont
  {El-Ganainy}}, \bibinfo {author} {\bibfnamefont {K.~G.}\ \bibnamefont
  {Makris}}, \bibinfo {author} {\bibfnamefont {M.}~\bibnamefont {Khajavikhan}},
  \bibinfo {author} {\bibfnamefont {Z.~H.}\ \bibnamefont {Musslimani}},
  \bibinfo {author} {\bibfnamefont {S.}~\bibnamefont {Rotter}}, \ and\ \bibinfo
  {author} {\bibfnamefont {D.~N.}\ \bibnamefont {Christodoulides}},\ }\href
  {\doibase 10.1038/nphys4323} {\bibfield  {journal} {\bibinfo  {journal}
  {Nature Physics}\ }\textbf {\bibinfo {volume} {14}},\ \bibinfo {pages} {11}
  (\bibinfo {year} {2018})}\BibitemShut {NoStop}%
\bibitem [{\citenamefont {El-Ganainy}\ \emph {et~al.}(2019)\citenamefont
  {El-Ganainy}, \citenamefont {Khajavikhan}, \citenamefont {Christodoulides},\
  and\ \citenamefont {Ozdemir}}]{El_Ganainy2019}%
  \BibitemOpen
  \bibfield  {author} {\bibinfo {author} {\bibfnamefont {R.}~\bibnamefont
  {El-Ganainy}}, \bibinfo {author} {\bibfnamefont {M.}~\bibnamefont
  {Khajavikhan}}, \bibinfo {author} {\bibfnamefont {D.~N.}\ \bibnamefont
  {Christodoulides}}, \ and\ \bibinfo {author} {\bibfnamefont {S.~K.}\
  \bibnamefont {Ozdemir}},\ }\href {\doibase 10.1038/s42005-019-0130-z}
  {\bibfield  {journal} {\bibinfo  {journal} {Communications Physics}\ }\textbf
  {\bibinfo {volume} {2}},\ \bibinfo {pages} {37} (\bibinfo {year}
  {2019})}\BibitemShut {NoStop}%
\bibitem [{\citenamefont {Ashida}\ \emph {et~al.}(2020)\citenamefont {Ashida},
  \citenamefont {Gong},\ and\ \citenamefont {Ueda}}]{Ashida_2020}%
  \BibitemOpen
  \bibfield  {author} {\bibinfo {author} {\bibfnamefont {Y.}~\bibnamefont
  {Ashida}}, \bibinfo {author} {\bibfnamefont {Z.}~\bibnamefont {Gong}}, \ and\
  \bibinfo {author} {\bibfnamefont {M.}~\bibnamefont {Ueda}},\ }\href {\doibase
  10.1080/00018732.2021.1876991} {\bibfield  {journal} {\bibinfo  {journal}
  {Advances in Physics}\ }\textbf {\bibinfo {volume} {69}},\ \bibinfo {pages}
  {249} (\bibinfo {year} {2020})}\BibitemShut {NoStop}%
\bibitem [{\citenamefont {B\"uttiker}(1986)}]{buttiker_paper}%
  \BibitemOpen
  \bibfield  {author} {\bibinfo {author} {\bibfnamefont {M.}~\bibnamefont
  {B\"uttiker}},\ }\href {\doibase 10.1103/PhysRevLett.57.1761} {\bibfield
  {journal} {\bibinfo  {journal} {Phys. Rev. Lett.}\ }\textbf {\bibinfo
  {volume} {57}},\ \bibinfo {pages} {1761} (\bibinfo {year}
  {1986})}\BibitemShut {NoStop}%
\bibitem [{\citenamefont {Maassen}\ \emph {et~al.}(2009)\citenamefont
  {Maassen}, \citenamefont {Zahid},\ and\ \citenamefont
  {Guo}}]{molecular_buttiker}%
  \BibitemOpen
  \bibfield  {author} {\bibinfo {author} {\bibfnamefont {J.}~\bibnamefont
  {Maassen}}, \bibinfo {author} {\bibfnamefont {F.}~\bibnamefont {Zahid}}, \
  and\ \bibinfo {author} {\bibfnamefont {H.}~\bibnamefont {Guo}},\ }\href
  {\doibase 10.1103/PhysRevB.80.125423} {\bibfield  {journal} {\bibinfo
  {journal} {Phys. Rev. B}\ }\textbf {\bibinfo {volume} {80}},\ \bibinfo
  {pages} {125423} (\bibinfo {year} {2009})}\BibitemShut {NoStop}%
\bibitem [{\citenamefont {Cattena}\ \emph {et~al.}(2010)\citenamefont
  {Cattena}, \citenamefont {Bustos-Mar\'un},\ and\ \citenamefont
  {Pastawski}}]{molecular_dephasing1}%
  \BibitemOpen
  \bibfield  {author} {\bibinfo {author} {\bibfnamefont {C.~J.}\ \bibnamefont
  {Cattena}}, \bibinfo {author} {\bibfnamefont {R.~A.}\ \bibnamefont
  {Bustos-Mar\'un}}, \ and\ \bibinfo {author} {\bibfnamefont {H.~M.}\
  \bibnamefont {Pastawski}},\ }\href {\doibase 10.1103/PhysRevB.82.144201}
  {\bibfield  {journal} {\bibinfo  {journal} {Phys. Rev. B}\ }\textbf {\bibinfo
  {volume} {82}},\ \bibinfo {pages} {144201} (\bibinfo {year}
  {2010})}\BibitemShut {NoStop}%
\bibitem [{\citenamefont {Nozaki}\ \emph {et~al.}(2012)\citenamefont {Nozaki},
  \citenamefont {Gomes~da Rocha}, \citenamefont {Pastawski},\ and\
  \citenamefont {Cuniberti}}]{molecular_dephasing2}%
  \BibitemOpen
  \bibfield  {author} {\bibinfo {author} {\bibfnamefont {D.}~\bibnamefont
  {Nozaki}}, \bibinfo {author} {\bibfnamefont {C.}~\bibnamefont {Gomes~da
  Rocha}}, \bibinfo {author} {\bibfnamefont {H.~M.}\ \bibnamefont {Pastawski}},
  \ and\ \bibinfo {author} {\bibfnamefont {G.}~\bibnamefont {Cuniberti}},\
  }\href {\doibase 10.1103/PhysRevB.85.155327} {\bibfield  {journal} {\bibinfo
  {journal} {Phys. Rev. B}\ }\textbf {\bibinfo {volume} {85}},\ \bibinfo
  {pages} {155327} (\bibinfo {year} {2012})}\BibitemShut {NoStop}%
\bibitem [{\citenamefont {Nozaki}\ \emph {et~al.}(2008)\citenamefont {Nozaki},
  \citenamefont {Girard},\ and\ \citenamefont
  {Yoshizawa}}]{molecular_dephasing4}%
  \BibitemOpen
  \bibfield  {author} {\bibinfo {author} {\bibfnamefont {D.}~\bibnamefont
  {Nozaki}}, \bibinfo {author} {\bibfnamefont {Y.}~\bibnamefont {Girard}}, \
  and\ \bibinfo {author} {\bibfnamefont {K.}~\bibnamefont {Yoshizawa}},\ }\href
  {\doibase 10.1021/jp806806j} {\bibfield  {journal} {\bibinfo  {journal} {The
  Journal of Physical Chemistry C}\ }\textbf {\bibinfo {volume} {112}},\
  \bibinfo {pages} {17408} (\bibinfo {year} {2008})}\BibitemShut {NoStop}%
\bibitem [{\citenamefont {Kilgour}\ and\ \citenamefont
  {Segal}(2016)}]{segal-probe-1}%
  \BibitemOpen
  \bibfield  {author} {\bibinfo {author} {\bibfnamefont {M.}~\bibnamefont
  {Kilgour}}\ and\ \bibinfo {author} {\bibfnamefont {D.}~\bibnamefont
  {Segal}},\ }\href {\doibase 10.1063/1.4944470} {\bibfield  {journal}
  {\bibinfo  {journal} {The Journal of Chemical Physics}\ }\textbf {\bibinfo
  {volume} {144}},\ \bibinfo {pages} {124107} (\bibinfo {year}
  {2016})}\BibitemShut {NoStop}%
\bibitem [{\citenamefont {Kilgour}\ and\ \citenamefont
  {Segal}(2015)}]{segal-probe-2}%
  \BibitemOpen
  \bibfield  {author} {\bibinfo {author} {\bibfnamefont {M.}~\bibnamefont
  {Kilgour}}\ and\ \bibinfo {author} {\bibfnamefont {D.}~\bibnamefont
  {Segal}},\ }\href {\doibase 10.1063/1.4926395} {\bibfield  {journal}
  {\bibinfo  {journal} {The Journal of Chemical Physics}\ }\textbf {\bibinfo
  {volume} {143}},\ \bibinfo {pages} {024111} (\bibinfo {year}
  {2015})}\BibitemShut {NoStop}%
\bibitem [{\citenamefont {Korol}\ \emph {et~al.}(2018)\citenamefont {Korol},
  \citenamefont {Kilgour},\ and\ \citenamefont {Segal}}]{segal-probe-3}%
  \BibitemOpen
  \bibfield  {author} {\bibinfo {author} {\bibfnamefont {R.}~\bibnamefont
  {Korol}}, \bibinfo {author} {\bibfnamefont {M.}~\bibnamefont {Kilgour}}, \
  and\ \bibinfo {author} {\bibfnamefont {D.}~\bibnamefont {Segal}},\ }\href
  {\doibase https://doi.org/10.1016/j.cpc.2017.10.005} {\bibfield  {journal}
  {\bibinfo  {journal} {Computer Physics Communications}\ }\textbf {\bibinfo
  {volume} {224}},\ \bibinfo {pages} {396} (\bibinfo {year}
  {2018})}\BibitemShut {NoStop}%
\bibitem [{\citenamefont {Bandyopadhyay}\ and\ \citenamefont
  {Segal}(2011)}]{malay}%
  \BibitemOpen
  \bibfield  {author} {\bibinfo {author} {\bibfnamefont {M.}~\bibnamefont
  {Bandyopadhyay}}\ and\ \bibinfo {author} {\bibfnamefont {D.}~\bibnamefont
  {Segal}},\ }\href {\doibase 10.1103/PhysRevE.84.011151} {\bibfield  {journal}
  {\bibinfo  {journal} {Phys. Rev. E}\ }\textbf {\bibinfo {volume} {84}},\
  \bibinfo {pages} {011151} (\bibinfo {year} {2011})}\BibitemShut {NoStop}%
\bibitem [{\citenamefont {Bedkihal}\ \emph {et~al.}(2013)\citenamefont
  {Bedkihal}, \citenamefont {Bandyopadhyay},\ and\ \citenamefont
  {Segal}}]{Bedkihal2013}%
  \BibitemOpen
  \bibfield  {author} {\bibinfo {author} {\bibfnamefont {S.}~\bibnamefont
  {Bedkihal}}, \bibinfo {author} {\bibfnamefont {M.}~\bibnamefont
  {Bandyopadhyay}}, \ and\ \bibinfo {author} {\bibfnamefont {D.}~\bibnamefont
  {Segal}},\ }\href {\doibase 10.1140/epjb/e2013-40971-7} {\bibfield  {journal}
  {\bibinfo  {journal} {The European Physical Journal B}\ }\textbf {\bibinfo
  {volume} {86}},\ \bibinfo {pages} {506} (\bibinfo {year} {2013})}\BibitemShut
  {NoStop}%
\bibitem [{\citenamefont {D'Amato}\ and\ \citenamefont
  {Pastawski}(1990)}]{Pastawski}%
  \BibitemOpen
  \bibfield  {author} {\bibinfo {author} {\bibfnamefont {J.~L.}\ \bibnamefont
  {D'Amato}}\ and\ \bibinfo {author} {\bibfnamefont {H.~M.}\ \bibnamefont
  {Pastawski}},\ }\href {\doibase 10.1103/PhysRevB.41.7411} {\bibfield
  {journal} {\bibinfo  {journal} {Phys. Rev. B}\ }\textbf {\bibinfo {volume}
  {41}},\ \bibinfo {pages} {7411} (\bibinfo {year} {1990})}\BibitemShut
  {NoStop}%
\bibitem [{\citenamefont {Roy}(2007)}]{Roy-1}%
  \BibitemOpen
  \bibfield  {author} {\bibinfo {author} {\bibfnamefont {D.}~\bibnamefont
  {Roy}},\ }\href {\doibase 10.1088/0953-8984/20/02/025206} {\bibfield
  {journal} {\bibinfo  {journal} {Journal of Physics: Condensed Matter}\
  }\textbf {\bibinfo {volume} {20}},\ \bibinfo {pages} {025206} (\bibinfo
  {year} {2007})}\BibitemShut {NoStop}%
\bibitem [{\citenamefont {Roy}\ and\ \citenamefont
  {Dhar}(2007)}]{self-2-Abhishek}%
  \BibitemOpen
  \bibfield  {author} {\bibinfo {author} {\bibfnamefont {D.}~\bibnamefont
  {Roy}}\ and\ \bibinfo {author} {\bibfnamefont {A.}~\bibnamefont {Dhar}},\
  }\href {\doibase 10.1103/PhysRevB.75.195110} {\bibfield  {journal} {\bibinfo
  {journal} {Phys. Rev. B}\ }\textbf {\bibinfo {volume} {75}},\ \bibinfo
  {pages} {195110} (\bibinfo {year} {2007})}\BibitemShut {NoStop}%
\bibitem [{\citenamefont {Zerah-Harush}\ and\ \citenamefont
  {Dubi}(2020)}]{dephasing_dubi}%
  \BibitemOpen
  \bibfield  {author} {\bibinfo {author} {\bibfnamefont {E.}~\bibnamefont
  {Zerah-Harush}}\ and\ \bibinfo {author} {\bibfnamefont {Y.}~\bibnamefont
  {Dubi}},\ }\href {\doibase 10.1103/PhysRevResearch.2.023294} {\bibfield
  {journal} {\bibinfo  {journal} {Phys. Rev. Research}\ }\textbf {\bibinfo
  {volume} {2}},\ \bibinfo {pages} {023294} (\bibinfo {year}
  {2020})}\BibitemShut {NoStop}%
\bibitem [{\citenamefont {Kulkarni}\ \emph {et~al.}(2013)\citenamefont
  {Kulkarni}, \citenamefont {Tiwari},\ and\ \citenamefont
  {Segal}}]{kulkarni2013full}%
  \BibitemOpen
  \bibfield  {author} {\bibinfo {author} {\bibfnamefont {M.}~\bibnamefont
  {Kulkarni}}, \bibinfo {author} {\bibfnamefont {K.~L.}\ \bibnamefont
  {Tiwari}}, \ and\ \bibinfo {author} {\bibfnamefont {D.}~\bibnamefont
  {Segal}},\ }\href {\doibase 10.1088/1367-2630/15/1/013014} {\bibfield
  {journal} {\bibinfo  {journal} {New Journal of Physics}\ }\textbf {\bibinfo
  {volume} {15}},\ \bibinfo {pages} {013014} (\bibinfo {year}
  {2013})}\BibitemShut {NoStop}%
\bibitem [{\citenamefont {Kulkarni}\ \emph {et~al.}(2012)\citenamefont
  {Kulkarni}, \citenamefont {Tiwari},\ and\ \citenamefont
  {Segal}}]{kulkarni2012towards}%
  \BibitemOpen
  \bibfield  {author} {\bibinfo {author} {\bibfnamefont {M.}~\bibnamefont
  {Kulkarni}}, \bibinfo {author} {\bibfnamefont {K.~L.}\ \bibnamefont
  {Tiwari}}, \ and\ \bibinfo {author} {\bibfnamefont {D.}~\bibnamefont
  {Segal}},\ }\href {\doibase 10.1103/PhysRevB.86.155424} {\bibfield  {journal}
  {\bibinfo  {journal} {Phys. Rev. B}\ }\textbf {\bibinfo {volume} {86}},\
  \bibinfo {pages} {155424} (\bibinfo {year} {2012})}\BibitemShut {NoStop}%
\bibitem [{\citenamefont {{Saha}}\ \emph {et~al.}(2022)\citenamefont {{Saha}},
  \citenamefont {{Agarwalla}}, \citenamefont {{Kulkarni}},\ and\ \citenamefont
  {{Purkayastha}}}]{saha2022universal}%
  \BibitemOpen
  \bibfield  {author} {\bibinfo {author} {\bibfnamefont {M.}~\bibnamefont
  {{Saha}}}, \bibinfo {author} {\bibfnamefont {B.~K.}\ \bibnamefont
  {{Agarwalla}}}, \bibinfo {author} {\bibfnamefont {M.}~\bibnamefont
  {{Kulkarni}}}, \ and\ \bibinfo {author} {\bibfnamefont {A.}~\bibnamefont
  {{Purkayastha}}},\ }\href@noop {} {\  (\bibinfo {year} {2022})},\ \Eprint
  {http://arxiv.org/abs/2205.02214} {arXiv:2205.02214 [quant-ph]} \BibitemShut
  {NoStop}%
\bibitem [{\citenamefont {Last}(1993)}]{lastY}%
  \BibitemOpen
  \bibfield  {author} {\bibinfo {author} {\bibfnamefont {Y.}~\bibnamefont
  {Last}},\ }\href {\doibase 10.1007/BF02096752} {\bibfield  {journal}
  {\bibinfo  {journal} {Communications in Mathematical Physics}\ }\textbf
  {\bibinfo {volume} {151}},\ \bibinfo {pages} {183} (\bibinfo {year}
  {1993})}\BibitemShut {NoStop}%
\bibitem [{\citenamefont {Molinari}(1997)}]{molinari1}%
  \BibitemOpen
  \bibfield  {author} {\bibinfo {author} {\bibfnamefont {L.}~\bibnamefont
  {Molinari}},\ }\href {\doibase 10.1088/0305-4470/30/3/021} {\bibfield
  {journal} {\bibinfo  {journal} {Journal of Physics A: Mathematical and
  General}\ }\textbf {\bibinfo {volume} {30}},\ \bibinfo {pages} {983}
  (\bibinfo {year} {1997})}\BibitemShut {NoStop}%
\bibitem [{\citenamefont {Molinari}(1998)}]{molinari2}%
  \BibitemOpen
  \bibfield  {author} {\bibinfo {author} {\bibfnamefont {L.}~\bibnamefont
  {Molinari}},\ }\href {\doibase 10.1088/0305-4470/31/42/014} {\bibfield
  {journal} {\bibinfo  {journal} {Journal of Physics A: Mathematical and
  General}\ }\textbf {\bibinfo {volume} {31}},\ \bibinfo {pages} {8553}
  (\bibinfo {year} {1998})}\BibitemShut {NoStop}%
\bibitem [{\citenamefont {Plenio}\ and\ \citenamefont
  {Huelga}(2008)}]{Plenio_2008}%
  \BibitemOpen
  \bibfield  {author} {\bibinfo {author} {\bibfnamefont {M.~B.}\ \bibnamefont
  {Plenio}}\ and\ \bibinfo {author} {\bibfnamefont {S.~F.}\ \bibnamefont
  {Huelga}},\ }\href {\doibase 10.1088/1367-2630/10/11/113019} {\bibfield
  {journal} {\bibinfo  {journal} {New Journal of Physics}\ }\textbf {\bibinfo
  {volume} {10}},\ \bibinfo {pages} {113019} (\bibinfo {year}
  {2008})}\BibitemShut {NoStop}%
\bibitem [{\citenamefont {Rebentrost}\ \emph {et~al.}(2009)\citenamefont
  {Rebentrost}, \citenamefont {Mohseni}, \citenamefont {Kassal}, \citenamefont
  {Lloyd},\ and\ \citenamefont {Aspuru-Guzik}}]{Rebentrost_2009}%
  \BibitemOpen
  \bibfield  {author} {\bibinfo {author} {\bibfnamefont {P.}~\bibnamefont
  {Rebentrost}}, \bibinfo {author} {\bibfnamefont {M.}~\bibnamefont {Mohseni}},
  \bibinfo {author} {\bibfnamefont {I.}~\bibnamefont {Kassal}}, \bibinfo
  {author} {\bibfnamefont {S.}~\bibnamefont {Lloyd}}, \ and\ \bibinfo {author}
  {\bibfnamefont {A.}~\bibnamefont {Aspuru-Guzik}},\ }\href {\doibase
  10.1088/1367-2630/11/3/033003} {\bibfield  {journal} {\bibinfo  {journal}
  {New Journal of Physics}\ }\textbf {\bibinfo {volume} {11}},\ \bibinfo
  {pages} {033003} (\bibinfo {year} {2009})}\BibitemShut {NoStop}%
\bibitem [{\citenamefont {Caruso}\ \emph {et~al.}(2009)\citenamefont {Caruso},
  \citenamefont {Chin}, \citenamefont {Datta}, \citenamefont {Huelga},\ and\
  \citenamefont {Plenio}}]{Caruso_2009}%
  \BibitemOpen
  \bibfield  {author} {\bibinfo {author} {\bibfnamefont {F.}~\bibnamefont
  {Caruso}}, \bibinfo {author} {\bibfnamefont {A.~W.}\ \bibnamefont {Chin}},
  \bibinfo {author} {\bibfnamefont {A.}~\bibnamefont {Datta}}, \bibinfo
  {author} {\bibfnamefont {S.~F.}\ \bibnamefont {Huelga}}, \ and\ \bibinfo
  {author} {\bibfnamefont {M.~B.}\ \bibnamefont {Plenio}},\ }\href {\doibase
  10.1063/1.3223548} {\bibfield  {journal} {\bibinfo  {journal} {The Journal of
  Chemical Physics}\ }\textbf {\bibinfo {volume} {131}} (\bibinfo {year}
  {2009}),\ 10.1063/1.3223548},\ \bibinfo {note} {105106},\ \Eprint
  {http://arxiv.org/abs/https://pubs.aip.org/aip/jcp/article-pdf/doi/10.1063/1.3223548/15931008/105106\_1\_online.pdf}
  {https://pubs.aip.org/aip/jcp/article-pdf/doi/10.1063/1.3223548/15931008/105106\_1\_online.pdf}
  \BibitemShut {NoStop}%
\bibitem [{\citenamefont {Sowa}\ \emph {et~al.}(2017)\citenamefont {Sowa},
  \citenamefont {Mol}, \citenamefont {Briggs},\ and\ \citenamefont
  {Gauger}}]{Sowa_2017}%
  \BibitemOpen
  \bibfield  {author} {\bibinfo {author} {\bibfnamefont {J.~K.}\ \bibnamefont
  {Sowa}}, \bibinfo {author} {\bibfnamefont {J.~A.}\ \bibnamefont {Mol}},
  \bibinfo {author} {\bibfnamefont {G.~A.~D.}\ \bibnamefont {Briggs}}, \ and\
  \bibinfo {author} {\bibfnamefont {E.~M.}\ \bibnamefont {Gauger}},\ }\href
  {\doibase 10.1039/C7CP06237K} {\bibfield  {journal} {\bibinfo  {journal}
  {Phys. Chem. Chem. Phys.}\ }\textbf {\bibinfo {volume} {19}},\ \bibinfo
  {pages} {29534} (\bibinfo {year} {2017})}\BibitemShut {NoStop}%
\bibitem [{\citenamefont {Zerah-Harush}\ and\ \citenamefont
  {Dubi}(2018)}]{Harush_2018}%
  \BibitemOpen
  \bibfield  {author} {\bibinfo {author} {\bibfnamefont {E.}~\bibnamefont
  {Zerah-Harush}}\ and\ \bibinfo {author} {\bibfnamefont {Y.}~\bibnamefont
  {Dubi}},\ }\href {\doibase 10.1021/acs.jpclett.7b03306} {\bibfield  {journal}
  {\bibinfo  {journal} {The Journal of Physical Chemistry Letters}\ }\textbf
  {\bibinfo {volume} {9}},\ \bibinfo {pages} {1689} (\bibinfo {year} {2018})},\
  \bibinfo {note} {pMID: 29537848}\BibitemShut {NoStop}%
\bibitem [{\citenamefont {Dutta}\ and\ \citenamefont
  {Bagchi}(2017)}]{Dutta_2017}%
  \BibitemOpen
  \bibfield  {author} {\bibinfo {author} {\bibfnamefont {R.}~\bibnamefont
  {Dutta}}\ and\ \bibinfo {author} {\bibfnamefont {B.}~\bibnamefont {Bagchi}},\
  }\href {\doibase 10.1021/acs.jpclett.7b02480} {\bibfield  {journal} {\bibinfo
   {journal} {The Journal of Physical Chemistry Letters}\ }\textbf {\bibinfo
  {volume} {8}},\ \bibinfo {pages} {5566} (\bibinfo {year} {2017})},\ \bibinfo
  {note} {pMID: 29083925}\BibitemShut {NoStop}%
\bibitem [{\citenamefont {Chiaracane}\ \emph {et~al.}(2022)\citenamefont
  {Chiaracane}, \citenamefont {Purkayastha}, \citenamefont {Mitchison},\ and\
  \citenamefont {Goold}}]{buttikerdephase}%
  \BibitemOpen
  \bibfield  {author} {\bibinfo {author} {\bibfnamefont {C.}~\bibnamefont
  {Chiaracane}}, \bibinfo {author} {\bibfnamefont {A.}~\bibnamefont
  {Purkayastha}}, \bibinfo {author} {\bibfnamefont {M.~T.}\ \bibnamefont
  {Mitchison}}, \ and\ \bibinfo {author} {\bibfnamefont {J.}~\bibnamefont
  {Goold}},\ }\href {\doibase 10.1103/PhysRevB.105.134203} {\bibfield
  {journal} {\bibinfo  {journal} {Phys. Rev. B}\ }\textbf {\bibinfo {volume}
  {105}},\ \bibinfo {pages} {134203} (\bibinfo {year} {2022})}\BibitemShut
  {NoStop}%
\bibitem [{\citenamefont {Lacerda}\ \emph {et~al.}(2021)\citenamefont
  {Lacerda}, \citenamefont {Goold},\ and\ \citenamefont
  {Landi}}]{Landi-latest}%
  \BibitemOpen
  \bibfield  {author} {\bibinfo {author} {\bibfnamefont {A.~M.}\ \bibnamefont
  {Lacerda}}, \bibinfo {author} {\bibfnamefont {J.}~\bibnamefont {Goold}}, \
  and\ \bibinfo {author} {\bibfnamefont {G.~T.}\ \bibnamefont {Landi}},\ }\href
  {\doibase 10.1103/PhysRevB.104.174203} {\bibfield  {journal} {\bibinfo
  {journal} {Phys. Rev. B}\ }\textbf {\bibinfo {volume} {104}},\ \bibinfo
  {pages} {174203} (\bibinfo {year} {2021})}\BibitemShut {NoStop}%
\bibitem [{\citenamefont {Dwiputra}\ and\ \citenamefont {Zen}(2021)}]{ENQT1}%
  \BibitemOpen
  \bibfield  {author} {\bibinfo {author} {\bibfnamefont {D.}~\bibnamefont
  {Dwiputra}}\ and\ \bibinfo {author} {\bibfnamefont {F.~P.}\ \bibnamefont
  {Zen}},\ }\href {\doibase 10.1103/PhysRevA.104.022205} {\bibfield  {journal}
  {\bibinfo  {journal} {Phys. Rev. A}\ }\textbf {\bibinfo {volume} {104}},\
  \bibinfo {pages} {022205} (\bibinfo {year} {2021})}\BibitemShut {NoStop}%
\bibitem [{\citenamefont {Turkeshi}\ and\ \citenamefont
  {Schir\'o}(2021)}]{diffusion1}%
  \BibitemOpen
  \bibfield  {author} {\bibinfo {author} {\bibfnamefont {X.}~\bibnamefont
  {Turkeshi}}\ and\ \bibinfo {author} {\bibfnamefont {M.}~\bibnamefont
  {Schir\'o}},\ }\href {\doibase 10.1103/PhysRevB.104.144301} {\bibfield
  {journal} {\bibinfo  {journal} {Phys. Rev. B}\ }\textbf {\bibinfo {volume}
  {104}},\ \bibinfo {pages} {144301} (\bibinfo {year} {2021})}\BibitemShut
  {NoStop}%
\bibitem [{\citenamefont {Saha}\ \emph {et~al.}(2022)\citenamefont {Saha},
  \citenamefont {Venkatesh},\ and\ \citenamefont {Agarwalla}}]{saha_buttiker}%
  \BibitemOpen
  \bibfield  {author} {\bibinfo {author} {\bibfnamefont {M.}~\bibnamefont
  {Saha}}, \bibinfo {author} {\bibfnamefont {B.~P.}\ \bibnamefont {Venkatesh}},
  \ and\ \bibinfo {author} {\bibfnamefont {B.~K.}\ \bibnamefont {Agarwalla}},\
  }\href {\doibase 10.1103/PhysRevB.105.224204} {\bibfield  {journal} {\bibinfo
   {journal} {Phys. Rev. B}\ }\textbf {\bibinfo {volume} {105}},\ \bibinfo
  {pages} {224204} (\bibinfo {year} {2022})}\BibitemShut {NoStop}%
\bibitem [{\citenamefont {Medvedyeva}\ \emph {et~al.}(2016)\citenamefont
  {Medvedyeva}, \citenamefont {Prosen},\ and\ \citenamefont {\ifmmode
  \check{Z}\else \v{Z}\fi{}nidari\ifmmode~\check{c}\else
  \v{c}\fi{}}}]{dephasing-interaction1}%
  \BibitemOpen
  \bibfield  {author} {\bibinfo {author} {\bibfnamefont {M.~V.}\ \bibnamefont
  {Medvedyeva}}, \bibinfo {author} {\bibfnamefont {T.~c.~v.}\ \bibnamefont
  {Prosen}}, \ and\ \bibinfo {author} {\bibfnamefont {M.}~\bibnamefont
  {\ifmmode \check{Z}\else \v{Z}\fi{}nidari\ifmmode~\check{c}\else
  \v{c}\fi{}}},\ }\href {\doibase 10.1103/PhysRevB.93.094205} {\bibfield
  {journal} {\bibinfo  {journal} {Phys. Rev. B}\ }\textbf {\bibinfo {volume}
  {93}},\ \bibinfo {pages} {094205} (\bibinfo {year} {2016})}\BibitemShut
  {NoStop}%
\bibitem [{\citenamefont {Žnidarič}\ \emph {et~al.}(2017)\citenamefont
  {Žnidarič}, \citenamefont {Mendoza-Arenas}, \citenamefont {Clark},\ and\
  \citenamefont {Goold}}]{dephasing_spin_interaction2}%
  \BibitemOpen
  \bibfield  {author} {\bibinfo {author} {\bibfnamefont {M.}~\bibnamefont
  {Žnidarič}}, \bibinfo {author} {\bibfnamefont {J.~J.}\ \bibnamefont
  {Mendoza-Arenas}}, \bibinfo {author} {\bibfnamefont {S.~R.}\ \bibnamefont
  {Clark}}, \ and\ \bibinfo {author} {\bibfnamefont {J.}~\bibnamefont
  {Goold}},\ }\href {\doibase https://doi.org/10.1002/andp.201600298}
  {\bibfield  {journal} {\bibinfo  {journal} {Annalen der Physik}\ }\textbf
  {\bibinfo {volume} {529}},\ \bibinfo {pages} {1600298} (\bibinfo {year}
  {2017})}\BibitemShut {NoStop}%
\bibitem [{\citenamefont {\ifmmode \check{Z}\else
  \v{Z}\fi{}nidari\ifmmode~\check{c}\else
  \v{c}\fi{}}(2018)}]{dephasing_interaction3}%
  \BibitemOpen
  \bibfield  {author} {\bibinfo {author} {\bibfnamefont {M.}~\bibnamefont
  {\ifmmode \check{Z}\else \v{Z}\fi{}nidari\ifmmode~\check{c}\else
  \v{c}\fi{}}},\ }\href {\doibase 10.1103/PhysRevB.97.214202} {\bibfield
  {journal} {\bibinfo  {journal} {Phys. Rev. B}\ }\textbf {\bibinfo {volume}
  {97}},\ \bibinfo {pages} {214202} (\bibinfo {year} {2018})}\BibitemShut
  {NoStop}%
\bibitem [{\citenamefont {Mendoza-Arenas}\ \emph {et~al.}(2013)\citenamefont
  {Mendoza-Arenas}, \citenamefont {Grujic}, \citenamefont {Jaksch},\ and\
  \citenamefont {Clark}}]{Mendoza-Arenas_2013}%
  \BibitemOpen
  \bibfield  {author} {\bibinfo {author} {\bibfnamefont {J.~J.}\ \bibnamefont
  {Mendoza-Arenas}}, \bibinfo {author} {\bibfnamefont {T.}~\bibnamefont
  {Grujic}}, \bibinfo {author} {\bibfnamefont {D.}~\bibnamefont {Jaksch}}, \
  and\ \bibinfo {author} {\bibfnamefont {S.~R.}\ \bibnamefont {Clark}},\ }\href
  {\doibase 10.1103/PhysRevB.87.235130} {\bibfield  {journal} {\bibinfo
  {journal} {Phys. Rev. B}\ }\textbf {\bibinfo {volume} {87}},\ \bibinfo
  {pages} {235130} (\bibinfo {year} {2013})}\BibitemShut {NoStop}%
\bibitem [{\citenamefont {Maier}\ \emph {et~al.}(2019)\citenamefont {Maier},
  \citenamefont {Brydges}, \citenamefont {Jurcevic}, \citenamefont {Trautmann},
  \citenamefont {Hempel}, \citenamefont {Lanyon}, \citenamefont {Hauke},
  \citenamefont {Blatt},\ and\ \citenamefont {Roos}}]{Maier_2019}%
  \BibitemOpen
  \bibfield  {author} {\bibinfo {author} {\bibfnamefont {C.}~\bibnamefont
  {Maier}}, \bibinfo {author} {\bibfnamefont {T.}~\bibnamefont {Brydges}},
  \bibinfo {author} {\bibfnamefont {P.}~\bibnamefont {Jurcevic}}, \bibinfo
  {author} {\bibfnamefont {N.}~\bibnamefont {Trautmann}}, \bibinfo {author}
  {\bibfnamefont {C.}~\bibnamefont {Hempel}}, \bibinfo {author} {\bibfnamefont
  {B.~P.}\ \bibnamefont {Lanyon}}, \bibinfo {author} {\bibfnamefont
  {P.}~\bibnamefont {Hauke}}, \bibinfo {author} {\bibfnamefont
  {R.}~\bibnamefont {Blatt}}, \ and\ \bibinfo {author} {\bibfnamefont {C.~F.}\
  \bibnamefont {Roos}},\ }\href {\doibase 10.1103/PhysRevLett.122.050501}
  {\bibfield  {journal} {\bibinfo  {journal} {Phys. Rev. Lett.}\ }\textbf
  {\bibinfo {volume} {122}},\ \bibinfo {pages} {050501} (\bibinfo {year}
  {2019})}\BibitemShut {NoStop}%
\bibitem [{\citenamefont {Le{\'o}n-Montiel}\ \emph {et~al.}(2015)\citenamefont
  {Le{\'o}n-Montiel}, \citenamefont {Quiroz-Ju{\'a}rez}, \citenamefont
  {Quintero-Torres}, \citenamefont {Dom{\'i}nguez-Ju{\'a}rez}, \citenamefont
  {Moya-Cessa}, \citenamefont {Torres},\ and\ \citenamefont
  {Arag{\'o}n}}]{Leon-Montiel_2015}%
  \BibitemOpen
  \bibfield  {author} {\bibinfo {author} {\bibfnamefont {R.~d.~J.}\
  \bibnamefont {Le{\'o}n-Montiel}}, \bibinfo {author} {\bibfnamefont {M.~A.}\
  \bibnamefont {Quiroz-Ju{\'a}rez}}, \bibinfo {author} {\bibfnamefont
  {R.}~\bibnamefont {Quintero-Torres}}, \bibinfo {author} {\bibfnamefont
  {J.~L.}\ \bibnamefont {Dom{\'i}nguez-Ju{\'a}rez}}, \bibinfo {author}
  {\bibfnamefont {H.~M.}\ \bibnamefont {Moya-Cessa}}, \bibinfo {author}
  {\bibfnamefont {J.~P.}\ \bibnamefont {Torres}}, \ and\ \bibinfo {author}
  {\bibfnamefont {J.~L.}\ \bibnamefont {Arag{\'o}n}},\ }\href {\doibase
  10.1038/srep17339} {\bibfield  {journal} {\bibinfo  {journal} {Scientific
  Reports}\ }\textbf {\bibinfo {volume} {5}},\ \bibinfo {pages} {17339}
  (\bibinfo {year} {2015})}\BibitemShut {NoStop}%
\bibitem [{\citenamefont {Biggerstaff}\ \emph {et~al.}(2016)\citenamefont
  {Biggerstaff}, \citenamefont {Heilmann}, \citenamefont {Zecevik},
  \citenamefont {Gr{\"a}fe}, \citenamefont {Broome}, \citenamefont {Fedrizzi},
  \citenamefont {Nolte}, \citenamefont {Szameit}, \citenamefont {White},\ and\
  \citenamefont {Kassal}}]{Biggerstaff_2016}%
  \BibitemOpen
  \bibfield  {author} {\bibinfo {author} {\bibfnamefont {D.~N.}\ \bibnamefont
  {Biggerstaff}}, \bibinfo {author} {\bibfnamefont {R.}~\bibnamefont
  {Heilmann}}, \bibinfo {author} {\bibfnamefont {A.~A.}\ \bibnamefont
  {Zecevik}}, \bibinfo {author} {\bibfnamefont {M.}~\bibnamefont {Gr{\"a}fe}},
  \bibinfo {author} {\bibfnamefont {M.~A.}\ \bibnamefont {Broome}}, \bibinfo
  {author} {\bibfnamefont {A.}~\bibnamefont {Fedrizzi}}, \bibinfo {author}
  {\bibfnamefont {S.}~\bibnamefont {Nolte}}, \bibinfo {author} {\bibfnamefont
  {A.}~\bibnamefont {Szameit}}, \bibinfo {author} {\bibfnamefont {A.~G.}\
  \bibnamefont {White}}, \ and\ \bibinfo {author} {\bibfnamefont
  {I.}~\bibnamefont {Kassal}},\ }\href {\doibase 10.1038/ncomms11282}
  {\bibfield  {journal} {\bibinfo  {journal} {Nature Communications}\ }\textbf
  {\bibinfo {volume} {7}},\ \bibinfo {pages} {11282} (\bibinfo {year}
  {2016})}\BibitemShut {NoStop}%
\bibitem [{\citenamefont {Viciani}\ \emph {et~al.}(2015)\citenamefont
  {Viciani}, \citenamefont {Lima}, \citenamefont {Bellini},\ and\ \citenamefont
  {Caruso}}]{Viciani_2015}%
  \BibitemOpen
  \bibfield  {author} {\bibinfo {author} {\bibfnamefont {S.}~\bibnamefont
  {Viciani}}, \bibinfo {author} {\bibfnamefont {M.}~\bibnamefont {Lima}},
  \bibinfo {author} {\bibfnamefont {M.}~\bibnamefont {Bellini}}, \ and\
  \bibinfo {author} {\bibfnamefont {F.}~\bibnamefont {Caruso}},\ }\href
  {\doibase 10.1103/PhysRevLett.115.083601} {\bibfield  {journal} {\bibinfo
  {journal} {Phys. Rev. Lett.}\ }\textbf {\bibinfo {volume} {115}},\ \bibinfo
  {pages} {083601} (\bibinfo {year} {2015})}\BibitemShut {NoStop}%
\bibitem [{\citenamefont {Harris}\ \emph {et~al.}(2017)\citenamefont {Harris},
  \citenamefont {Steinbrecher}, \citenamefont {Prabhu}, \citenamefont {Lahini},
  \citenamefont {Mower}, \citenamefont {Bunandar}, \citenamefont {Chen},
  \citenamefont {Wong}, \citenamefont {Baehr-Jones}, \citenamefont {Hochberg},
  \citenamefont {Lloyd},\ and\ \citenamefont {Englund}}]{Harris_2017}%
  \BibitemOpen
  \bibfield  {author} {\bibinfo {author} {\bibfnamefont {N.~C.}\ \bibnamefont
  {Harris}}, \bibinfo {author} {\bibfnamefont {G.~R.}\ \bibnamefont
  {Steinbrecher}}, \bibinfo {author} {\bibfnamefont {M.}~\bibnamefont
  {Prabhu}}, \bibinfo {author} {\bibfnamefont {Y.}~\bibnamefont {Lahini}},
  \bibinfo {author} {\bibfnamefont {J.}~\bibnamefont {Mower}}, \bibinfo
  {author} {\bibfnamefont {D.}~\bibnamefont {Bunandar}}, \bibinfo {author}
  {\bibfnamefont {C.}~\bibnamefont {Chen}}, \bibinfo {author} {\bibfnamefont
  {F.~N.~C.}\ \bibnamefont {Wong}}, \bibinfo {author} {\bibfnamefont
  {T.}~\bibnamefont {Baehr-Jones}}, \bibinfo {author} {\bibfnamefont
  {M.}~\bibnamefont {Hochberg}}, \bibinfo {author} {\bibfnamefont
  {S.}~\bibnamefont {Lloyd}}, \ and\ \bibinfo {author} {\bibfnamefont
  {D.}~\bibnamefont {Englund}},\ }\href {\doibase 10.1038/nphoton.2017.95}
  {\bibfield  {journal} {\bibinfo  {journal} {Nature Photonics}\ }\textbf
  {\bibinfo {volume} {11}},\ \bibinfo {pages} {447} (\bibinfo {year}
  {2017})}\BibitemShut {NoStop}%
\bibitem [{\citenamefont {Krishna~Kumar}\ \emph {et~al.}(2017)\citenamefont
  {Krishna~Kumar}, \citenamefont {Bandurin}, \citenamefont {Pellegrino},
  \citenamefont {Cao}, \citenamefont {Principi}, \citenamefont {Guo},
  \citenamefont {Auton}, \citenamefont {Ben~Shalom}, \citenamefont
  {Ponomarenko}, \citenamefont {Falkovich}, \citenamefont {Watanabe},
  \citenamefont {Taniguchi}, \citenamefont {Grigorieva}, \citenamefont
  {Levitov}, \citenamefont {Polini},\ and\ \citenamefont
  {Geim}}]{superballistic1}%
  \BibitemOpen
  \bibfield  {author} {\bibinfo {author} {\bibfnamefont {R.}~\bibnamefont
  {Krishna~Kumar}}, \bibinfo {author} {\bibfnamefont {D.~A.}\ \bibnamefont
  {Bandurin}}, \bibinfo {author} {\bibfnamefont {F.~M.~D.}\ \bibnamefont
  {Pellegrino}}, \bibinfo {author} {\bibfnamefont {Y.}~\bibnamefont {Cao}},
  \bibinfo {author} {\bibfnamefont {A.}~\bibnamefont {Principi}}, \bibinfo
  {author} {\bibfnamefont {H.}~\bibnamefont {Guo}}, \bibinfo {author}
  {\bibfnamefont {G.~H.}\ \bibnamefont {Auton}}, \bibinfo {author}
  {\bibfnamefont {M.}~\bibnamefont {Ben~Shalom}}, \bibinfo {author}
  {\bibfnamefont {L.~A.}\ \bibnamefont {Ponomarenko}}, \bibinfo {author}
  {\bibfnamefont {G.}~\bibnamefont {Falkovich}}, \bibinfo {author}
  {\bibfnamefont {K.}~\bibnamefont {Watanabe}}, \bibinfo {author}
  {\bibfnamefont {T.}~\bibnamefont {Taniguchi}}, \bibinfo {author}
  {\bibfnamefont {I.~V.}\ \bibnamefont {Grigorieva}}, \bibinfo {author}
  {\bibfnamefont {L.~S.}\ \bibnamefont {Levitov}}, \bibinfo {author}
  {\bibfnamefont {M.}~\bibnamefont {Polini}}, \ and\ \bibinfo {author}
  {\bibfnamefont {A.~K.}\ \bibnamefont {Geim}},\ }\href {\doibase
  10.1038/nphys4240} {\bibfield  {journal} {\bibinfo  {journal} {Nature
  Physics}\ }\textbf {\bibinfo {volume} {13}},\ \bibinfo {pages} {1182}
  (\bibinfo {year} {2017})}\BibitemShut {NoStop}%
\bibitem [{\citenamefont {{Raichev}}(2022)}]{superballistic2}%
  \BibitemOpen
  \bibfield  {author} {\bibinfo {author} {\bibfnamefont {O.~E.}\ \bibnamefont
  {{Raichev}}},\ }\href@noop {} {\  (\bibinfo {year} {2022})},\ \Eprint
  {http://arxiv.org/abs/2202.06623} {arXiv:2202.06623 [cond-mat.mes-hall]}
  \BibitemShut {NoStop}%
\bibitem [{\citenamefont {Ginzburg}\ \emph {et~al.}(2021)\citenamefont
  {Ginzburg}, \citenamefont {Gold}, \citenamefont {R\"o\"osli}, \citenamefont
  {Reichl}, \citenamefont {Berl}, \citenamefont {Wegscheider}, \citenamefont
  {Ihn},\ and\ \citenamefont {Ensslin}}]{superballistic3}%
  \BibitemOpen
  \bibfield  {author} {\bibinfo {author} {\bibfnamefont {L.~V.}\ \bibnamefont
  {Ginzburg}}, \bibinfo {author} {\bibfnamefont {C.}~\bibnamefont {Gold}},
  \bibinfo {author} {\bibfnamefont {M.~P.}\ \bibnamefont {R\"o\"osli}},
  \bibinfo {author} {\bibfnamefont {C.}~\bibnamefont {Reichl}}, \bibinfo
  {author} {\bibfnamefont {M.}~\bibnamefont {Berl}}, \bibinfo {author}
  {\bibfnamefont {W.}~\bibnamefont {Wegscheider}}, \bibinfo {author}
  {\bibfnamefont {T.}~\bibnamefont {Ihn}}, \ and\ \bibinfo {author}
  {\bibfnamefont {K.}~\bibnamefont {Ensslin}},\ }\href {\doibase
  10.1103/PhysRevResearch.3.023033} {\bibfield  {journal} {\bibinfo  {journal}
  {Phys. Rev. Research}\ }\textbf {\bibinfo {volume} {3}},\ \bibinfo {pages}
  {023033} (\bibinfo {year} {2021})}\BibitemShut {NoStop}%
\bibitem [{\citenamefont {Szameit}\ \emph {et~al.}(2012)\citenamefont
  {Szameit}, \citenamefont {St\"{u}tzer}, \citenamefont {Kottos}, \citenamefont
  {T\"{u}nnermann}, \citenamefont {Nolte},\ and\ \citenamefont
  {Christodoulides}}]{superballistic4}%
  \BibitemOpen
  \bibfield  {author} {\bibinfo {author} {\bibfnamefont {A.}~\bibnamefont
  {Szameit}}, \bibinfo {author} {\bibfnamefont {S.}~\bibnamefont
  {St\"{u}tzer}}, \bibinfo {author} {\bibfnamefont {T.}~\bibnamefont {Kottos}},
  \bibinfo {author} {\bibfnamefont {A.}~\bibnamefont {T\"{u}nnermann}},
  \bibinfo {author} {\bibfnamefont {S.}~\bibnamefont {Nolte}}, \ and\ \bibinfo
  {author} {\bibfnamefont {D.~N.}\ \bibnamefont {Christodoulides}},\ }\href
  {http://opg.optica.org/abstract.cfm?URI=FiO-2012-FTh2G.1} {\bibfield
  {journal} {\bibinfo  {journal} {Frontiers in Optics 2012/Laser Science
  XXVIII}\ ,\ \bibinfo {pages} {FTh2G.1}} (\bibinfo {year} {2012})}\BibitemShut
  {NoStop}%
\bibitem [{\citenamefont {St\"{u}tzer}\ \emph {et~al.}(2013)\citenamefont
  {St\"{u}tzer}, \citenamefont {Kottos}, \citenamefont {T\"{u}nnermann},
  \citenamefont {Nolte}, \citenamefont {Christodoulides},\ and\ \citenamefont
  {Szameit}}]{Superballistic5}%
  \BibitemOpen
  \bibfield  {author} {\bibinfo {author} {\bibfnamefont {S.}~\bibnamefont
  {St\"{u}tzer}}, \bibinfo {author} {\bibfnamefont {T.}~\bibnamefont {Kottos}},
  \bibinfo {author} {\bibfnamefont {A.}~\bibnamefont {T\"{u}nnermann}},
  \bibinfo {author} {\bibfnamefont {S.}~\bibnamefont {Nolte}}, \bibinfo
  {author} {\bibfnamefont {D.~N.}\ \bibnamefont {Christodoulides}}, \ and\
  \bibinfo {author} {\bibfnamefont {A.}~\bibnamefont {Szameit}},\ }\href
  {http://opg.optica.org/abstract.cfm?URI=CLEO_Europe-2013-CK_8_3} {\bibfield
  {journal} {\bibinfo  {journal} {2013 Conference on Lasers and Electro-Optics
  - International Quantum Electronics Conference}\ } (\bibinfo {year}
  {2013})}\BibitemShut {NoStop}%
\bibitem [{\citenamefont {St\"{u}tzer}\ \emph {et~al.}(2012)\citenamefont
  {St\"{u}tzer}, \citenamefont {Kottos}, \citenamefont {T\"{u}nnermann},
  \citenamefont {Nolte}, \citenamefont {Christodoulides},\ and\ \citenamefont
  {Szameit}}]{Superballistic6}%
  \BibitemOpen
  \bibfield  {author} {\bibinfo {author} {\bibfnamefont {S.}~\bibnamefont
  {St\"{u}tzer}}, \bibinfo {author} {\bibfnamefont {T.}~\bibnamefont {Kottos}},
  \bibinfo {author} {\bibfnamefont {A.}~\bibnamefont {T\"{u}nnermann}},
  \bibinfo {author} {\bibfnamefont {S.}~\bibnamefont {Nolte}}, \bibinfo
  {author} {\bibfnamefont {D.~N.}\ \bibnamefont {Christodoulides}}, \ and\
  \bibinfo {author} {\bibfnamefont {A.}~\bibnamefont {Szameit}},\ }\href
  {http://opg.optica.org/abstract.cfm?URI=QELS-2012-QF1H.6} {\bibfield
  {journal} {\bibinfo  {journal} {Conference on Lasers and Electro-Optics
  2012}\ } (\bibinfo {year} {2012})}\BibitemShut {NoStop}%
\bibitem [{SM()}]{SM}%
  \BibitemOpen
  \href@noop {} {\bibinfo  {journal} {Supplemental Material}\ }\BibitemShut
  {NoStop}%
\bibitem [{\citenamefont {Sergueev}\ \emph {et~al.}(2005)\citenamefont
  {Sergueev}, \citenamefont {Roubtsov},\ and\ \citenamefont
  {Guo}}]{electron-phonon}%
  \BibitemOpen
\bibfield  {journal} {  }\bibfield  {author} {\bibinfo {author} {\bibfnamefont
  {N.}~\bibnamefont {Sergueev}}, \bibinfo {author} {\bibfnamefont
  {D.}~\bibnamefont {Roubtsov}}, \ and\ \bibinfo {author} {\bibfnamefont
  {H.}~\bibnamefont {Guo}},\ }\href {\doibase 10.1103/PhysRevLett.95.146803}
  {\bibfield  {journal} {\bibinfo  {journal} {Phys. Rev. Lett.}\ }\textbf
  {\bibinfo {volume} {95}},\ \bibinfo {pages} {146803} (\bibinfo {year}
  {2005})}\BibitemShut {NoStop}%
\bibitem [{\citenamefont {Pastawski}\ \emph {et~al.}(2002)\citenamefont
  {Pastawski}, \citenamefont {{Foa Torres}},\ and\ \citenamefont
  {Medina}}]{electron_phonon2}%
  \BibitemOpen
  \bibfield  {author} {\bibinfo {author} {\bibfnamefont {H.~M.}\ \bibnamefont
  {Pastawski}}, \bibinfo {author} {\bibfnamefont {L.}~\bibnamefont {{Foa
  Torres}}}, \ and\ \bibinfo {author} {\bibfnamefont {E.}~\bibnamefont
  {Medina}},\ }\href {\doibase https://doi.org/10.1016/S0301-0104(02)00565-7}
  {\bibfield  {journal} {\bibinfo  {journal} {Chemical Physics}\ }\textbf
  {\bibinfo {volume} {281}},\ \bibinfo {pages} {257} (\bibinfo {year}
  {2002})}\BibitemShut {NoStop}%
\bibitem [{\citenamefont {Brandbyge}\ \emph {et~al.}(2002)\citenamefont
  {Brandbyge}, \citenamefont {Mozos}, \citenamefont {Ordej\'on}, \citenamefont
  {Taylor},\ and\ \citenamefont {Stokbro}}]{Brandbyge_2002}%
  \BibitemOpen
  \bibfield  {author} {\bibinfo {author} {\bibfnamefont {M.}~\bibnamefont
  {Brandbyge}}, \bibinfo {author} {\bibfnamefont {J.-L.}\ \bibnamefont
  {Mozos}}, \bibinfo {author} {\bibfnamefont {P.}~\bibnamefont {Ordej\'on}},
  \bibinfo {author} {\bibfnamefont {J.}~\bibnamefont {Taylor}}, \ and\ \bibinfo
  {author} {\bibfnamefont {K.}~\bibnamefont {Stokbro}},\ }\href {\doibase
  10.1103/PhysRevB.65.165401} {\bibfield  {journal} {\bibinfo  {journal} {Phys.
  Rev. B}\ }\textbf {\bibinfo {volume} {65}},\ \bibinfo {pages} {165401}
  (\bibinfo {year} {2002})}\BibitemShut {NoStop}%
\bibitem [{\citenamefont {Nitzan}\ and\ \citenamefont
  {Ratner}(2003)}]{Nitzan_2003}%
  \BibitemOpen
  \bibfield  {author} {\bibinfo {author} {\bibfnamefont {A.}~\bibnamefont
  {Nitzan}}\ and\ \bibinfo {author} {\bibfnamefont {M.~A.}\ \bibnamefont
  {Ratner}},\ }\href {\doibase 10.1126/science.1081572} {\bibfield  {journal}
  {\bibinfo  {journal} {Science}\ }\textbf {\bibinfo {volume} {300}},\ \bibinfo
  {pages} {1384} (\bibinfo {year} {2003})},\ \Eprint
  {http://arxiv.org/abs/https://www.science.org/doi/pdf/10.1126/science.1081572}
  {https://www.science.org/doi/pdf/10.1126/science.1081572} \BibitemShut
  {NoStop}%
\bibitem [{\citenamefont {Larsen}\ \emph {et~al.}(2017)\citenamefont {Larsen},
  \citenamefont {Mortensen}, \citenamefont {Blomqvist}, \citenamefont
  {Castelli}, \citenamefont {Christensen}, \citenamefont {Dułak},
  \citenamefont {Friis}, \citenamefont {Groves}, \citenamefont {Hammer},
  \citenamefont {Hargus}, \citenamefont {Hermes}, \citenamefont {Jennings},
  \citenamefont {Jensen}, \citenamefont {Kermode}, \citenamefont {Kitchin},
  \citenamefont {Kolsbjerg}, \citenamefont {Kubal}, \citenamefont {Kaasbjerg},
  \citenamefont {Lysgaard}, \citenamefont {Maronsson}, \citenamefont {Maxson},
  \citenamefont {Olsen}, \citenamefont {Pastewka}, \citenamefont {Peterson},
  \citenamefont {Rostgaard}, \citenamefont {Schiøtz}, \citenamefont {Schütt},
  \citenamefont {Strange}, \citenamefont {Thygesen}, \citenamefont {Vegge},
  \citenamefont {Vilhelmsen}, \citenamefont {Walter}, \citenamefont {Zeng},\
  and\ \citenamefont {Jacobsen}}]{Hjorth_2017}%
  \BibitemOpen
  \bibfield  {author} {\bibinfo {author} {\bibfnamefont {A.~H.}\ \bibnamefont
  {Larsen}}, \bibinfo {author} {\bibfnamefont {J.~J.}\ \bibnamefont
  {Mortensen}}, \bibinfo {author} {\bibfnamefont {J.}~\bibnamefont
  {Blomqvist}}, \bibinfo {author} {\bibfnamefont {I.~E.}\ \bibnamefont
  {Castelli}}, \bibinfo {author} {\bibfnamefont {R.}~\bibnamefont
  {Christensen}}, \bibinfo {author} {\bibfnamefont {M.}~\bibnamefont {Dułak}},
  \bibinfo {author} {\bibfnamefont {J.}~\bibnamefont {Friis}}, \bibinfo
  {author} {\bibfnamefont {M.~N.}\ \bibnamefont {Groves}}, \bibinfo {author}
  {\bibfnamefont {B.}~\bibnamefont {Hammer}}, \bibinfo {author} {\bibfnamefont
  {C.}~\bibnamefont {Hargus}}, \bibinfo {author} {\bibfnamefont {E.~D.}\
  \bibnamefont {Hermes}}, \bibinfo {author} {\bibfnamefont {P.~C.}\
  \bibnamefont {Jennings}}, \bibinfo {author} {\bibfnamefont {P.~B.}\
  \bibnamefont {Jensen}}, \bibinfo {author} {\bibfnamefont {J.}~\bibnamefont
  {Kermode}}, \bibinfo {author} {\bibfnamefont {J.~R.}\ \bibnamefont
  {Kitchin}}, \bibinfo {author} {\bibfnamefont {E.~L.}\ \bibnamefont
  {Kolsbjerg}}, \bibinfo {author} {\bibfnamefont {J.}~\bibnamefont {Kubal}},
  \bibinfo {author} {\bibfnamefont {K.}~\bibnamefont {Kaasbjerg}}, \bibinfo
  {author} {\bibfnamefont {S.}~\bibnamefont {Lysgaard}}, \bibinfo {author}
  {\bibfnamefont {J.~B.}\ \bibnamefont {Maronsson}}, \bibinfo {author}
  {\bibfnamefont {T.}~\bibnamefont {Maxson}}, \bibinfo {author} {\bibfnamefont
  {T.}~\bibnamefont {Olsen}}, \bibinfo {author} {\bibfnamefont
  {L.}~\bibnamefont {Pastewka}}, \bibinfo {author} {\bibfnamefont
  {A.}~\bibnamefont {Peterson}}, \bibinfo {author} {\bibfnamefont
  {C.}~\bibnamefont {Rostgaard}}, \bibinfo {author} {\bibfnamefont
  {J.}~\bibnamefont {Schiøtz}}, \bibinfo {author} {\bibfnamefont
  {O.}~\bibnamefont {Schütt}}, \bibinfo {author} {\bibfnamefont
  {M.}~\bibnamefont {Strange}}, \bibinfo {author} {\bibfnamefont {K.~S.}\
  \bibnamefont {Thygesen}}, \bibinfo {author} {\bibfnamefont {T.}~\bibnamefont
  {Vegge}}, \bibinfo {author} {\bibfnamefont {L.}~\bibnamefont {Vilhelmsen}},
  \bibinfo {author} {\bibfnamefont {M.}~\bibnamefont {Walter}}, \bibinfo
  {author} {\bibfnamefont {Z.}~\bibnamefont {Zeng}}, \ and\ \bibinfo {author}
  {\bibfnamefont {K.~W.}\ \bibnamefont {Jacobsen}},\ }\href {\doibase
  10.1088/1361-648X/aa680e} {\bibfield  {journal} {\bibinfo  {journal} {Journal
  of Physics: Condensed Matter}\ }\textbf {\bibinfo {volume} {29}},\ \bibinfo
  {pages} {273002} (\bibinfo {year} {2017})}\BibitemShut {NoStop}%
\bibitem [{\citenamefont {Kurth}\ \emph {et~al.}(2019)\citenamefont {Kurth},
  \citenamefont {Jacob}, \citenamefont {Sobrino},\ and\ \citenamefont
  {Stefanucci}}]{Kurth_2019}%
  \BibitemOpen
  \bibfield  {author} {\bibinfo {author} {\bibfnamefont {S.}~\bibnamefont
  {Kurth}}, \bibinfo {author} {\bibfnamefont {D.}~\bibnamefont {Jacob}},
  \bibinfo {author} {\bibfnamefont {N.}~\bibnamefont {Sobrino}}, \ and\
  \bibinfo {author} {\bibfnamefont {G.}~\bibnamefont {Stefanucci}},\ }\href
  {\doibase 10.1103/PhysRevB.100.085114} {\bibfield  {journal} {\bibinfo
  {journal} {Phys. Rev. B}\ }\textbf {\bibinfo {volume} {100}},\ \bibinfo
  {pages} {085114} (\bibinfo {year} {2019})}\BibitemShut {NoStop}%
\bibitem [{\citenamefont {Smidstrup}\ \emph {et~al.}(2019)\citenamefont
  {Smidstrup}, \citenamefont {Markussen}, \citenamefont {Vancraeyveld},
  \citenamefont {Wellendorff}, \citenamefont {Schneider}, \citenamefont
  {Gunst}, \citenamefont {Verstichel}, \citenamefont {Stradi}, \citenamefont
  {Khomyakov}, \citenamefont {Vej-Hansen}, \citenamefont {Lee}, \citenamefont
  {Chill}, \citenamefont {Rasmussen}, \citenamefont {Penazzi}, \citenamefont
  {Corsetti}, \citenamefont {Ojanperä}, \citenamefont {Jensen}, \citenamefont
  {Palsgaard}, \citenamefont {Martinez}, \citenamefont {Blom}, \citenamefont
  {Brandbyge},\ and\ \citenamefont {Stokbro}}]{Smidstrup_2020}%
  \BibitemOpen
  \bibfield  {author} {\bibinfo {author} {\bibfnamefont {S.}~\bibnamefont
  {Smidstrup}}, \bibinfo {author} {\bibfnamefont {T.}~\bibnamefont
  {Markussen}}, \bibinfo {author} {\bibfnamefont {P.}~\bibnamefont
  {Vancraeyveld}}, \bibinfo {author} {\bibfnamefont {J.}~\bibnamefont
  {Wellendorff}}, \bibinfo {author} {\bibfnamefont {J.}~\bibnamefont
  {Schneider}}, \bibinfo {author} {\bibfnamefont {T.}~\bibnamefont {Gunst}},
  \bibinfo {author} {\bibfnamefont {B.}~\bibnamefont {Verstichel}}, \bibinfo
  {author} {\bibfnamefont {D.}~\bibnamefont {Stradi}}, \bibinfo {author}
  {\bibfnamefont {P.~A.}\ \bibnamefont {Khomyakov}}, \bibinfo {author}
  {\bibfnamefont {U.~G.}\ \bibnamefont {Vej-Hansen}}, \bibinfo {author}
  {\bibfnamefont {M.-E.}\ \bibnamefont {Lee}}, \bibinfo {author} {\bibfnamefont
  {S.~T.}\ \bibnamefont {Chill}}, \bibinfo {author} {\bibfnamefont
  {F.}~\bibnamefont {Rasmussen}}, \bibinfo {author} {\bibfnamefont
  {G.}~\bibnamefont {Penazzi}}, \bibinfo {author} {\bibfnamefont
  {F.}~\bibnamefont {Corsetti}}, \bibinfo {author} {\bibfnamefont
  {A.}~\bibnamefont {Ojanperä}}, \bibinfo {author} {\bibfnamefont
  {K.}~\bibnamefont {Jensen}}, \bibinfo {author} {\bibfnamefont {M.~L.~N.}\
  \bibnamefont {Palsgaard}}, \bibinfo {author} {\bibfnamefont {U.}~\bibnamefont
  {Martinez}}, \bibinfo {author} {\bibfnamefont {A.}~\bibnamefont {Blom}},
  \bibinfo {author} {\bibfnamefont {M.}~\bibnamefont {Brandbyge}}, \ and\
  \bibinfo {author} {\bibfnamefont {K.}~\bibnamefont {Stokbro}},\ }\href
  {\doibase 10.1088/1361-648X/ab4007} {\bibfield  {journal} {\bibinfo
  {journal} {Journal of Physics: Condensed Matter}\ }\textbf {\bibinfo {volume}
  {32}},\ \bibinfo {pages} {015901} (\bibinfo {year} {2019})}\BibitemShut
  {NoStop}%
\bibitem [{\citenamefont {Kou}\ \emph {et~al.}(2017)\citenamefont {Kou},
  \citenamefont {Ma}, \citenamefont {Sun}, \citenamefont {Heine},\ and\
  \citenamefont {Chen}}]{Kou_2017}%
  \BibitemOpen
  \bibfield  {author} {\bibinfo {author} {\bibfnamefont {L.}~\bibnamefont
  {Kou}}, \bibinfo {author} {\bibfnamefont {Y.}~\bibnamefont {Ma}}, \bibinfo
  {author} {\bibfnamefont {Z.}~\bibnamefont {Sun}}, \bibinfo {author}
  {\bibfnamefont {T.}~\bibnamefont {Heine}}, \ and\ \bibinfo {author}
  {\bibfnamefont {C.}~\bibnamefont {Chen}},\ }\href {\doibase
  10.1021/acs.jpclett.7b00222} {\bibfield  {journal} {\bibinfo  {journal} {The
  Journal of Physical Chemistry Letters}\ }\textbf {\bibinfo {volume} {8}},\
  \bibinfo {pages} {1905} (\bibinfo {year} {2017})},\ \bibinfo {note} {pMID:
  28394616}\BibitemShut {NoStop}%
\bibitem [{\citenamefont {Hasan}\ and\ \citenamefont
  {Kane}(2010)}]{Hasan_2010}%
  \BibitemOpen
  \bibfield  {author} {\bibinfo {author} {\bibfnamefont {M.~Z.}\ \bibnamefont
  {Hasan}}\ and\ \bibinfo {author} {\bibfnamefont {C.~L.}\ \bibnamefont
  {Kane}},\ }\href {\doibase 10.1103/RevModPhys.82.3045} {\bibfield  {journal}
  {\bibinfo  {journal} {Rev. Mod. Phys.}\ }\textbf {\bibinfo {volume} {82}},\
  \bibinfo {pages} {3045} (\bibinfo {year} {2010})}\BibitemShut {NoStop}%
\bibitem [{\citenamefont {Haug}\ and\ \citenamefont {Jauho}(2008)}]{Keldysh4}%
  \BibitemOpen
  \bibfield  {author} {\bibinfo {author} {\bibfnamefont {H.}~\bibnamefont
  {Haug}}\ and\ \bibinfo {author} {\bibfnamefont {A.}~\bibnamefont {Jauho}},\
  }\href@noop {} {\emph {\bibinfo {title} {Quantum Kinetics in Transport and
  Optics of Semiconductors}}}\ (\bibinfo {year} {Springer, New York,
  2008})\BibitemShut {NoStop}%
\end{thebibliography}%

\clearpage
\newpage

\section*{Supplementary material}
\setcounter{equation}{0}

\setcounter{figure}{0}

\renewcommand{\theequation}{S\arabic{equation}}

\renewcommand{\thefigure}{S\arabic{figure}}
\renewcommand{\thesection}{S\arabic{section}}

\section{B\"uttiker Voltage probe}
\label{NEGF}
In this section, we lay out some details about B\"uttiker voltage probe and its specific implementation in our case. Our entire setup that includes the finite tight-binding chain, the left/right reservoirs and the probes, are all of bilinear type. As a consequence, the non-equilibrium steady state electronic current (NESS) flowing out of each terminal follows the Landauer formula \cite{Keldysh4},
\begin{align}
\label{probe_current}
\mathcal{I}_{\nu}= \sum_{\alpha} \int_{-\infty}^{\infty} \mathcal{T}_{\nu \alpha}(\omega) \, \Big(f_{\nu} (\omega)-f_{\alpha}(\omega) \Big) \, d\omega.
\end{align}
Here $f_{\nu}(\omega)=(1+e^{\beta(\omega-\mu_{\nu})})^{-1}$ is the Fermi distribution function of $\nu$-th terminal with inverse temperature $\beta$ and chemical potential $\mu_{\nu}$ (same for $f_{\alpha}(\omega)$ with chemical potential $\mu_{\alpha})$. $\mathcal{T}_{\nu \alpha}(\omega)= {\rm Tr}[\Gamma_{\nu}(\omega) G^r(\omega) \Gamma_{\alpha}(\omega) G^a(\omega)]$ is the transmission probability for electrons to flow from $\alpha$-th terminal to $\nu$-th terminal. Here $G^r(\omega)$ is the retarded Green's function of the lattice chain in presence of the reservoirs. For sake of brevity, we suppress the superscript $r$ and denote the retarded Green's function as $G(\omega)$ given by 
\begin{equation}
G(\omega)=\big[\omega \mathbb{I}-H_C-\Sigma_L(\omega)-\Sigma_R(\omega)-\sum_{n=1}^{N} \Sigma_{P_n}(\omega)\big]^{-1}.
\label{Gr}
\end{equation} 
Here $\mathbb{I}$ being the $N \times N$ identity matrix and $H_C$ represents the single particle Hamiltonian corresponding to $\hat{H}_C$. The advanced Green's function is given by $G^a(\omega)=G^{\dagger}(\omega)$. $\Sigma_L(\omega), \Sigma_R(\omega)$ are the self-energy matrices for the left and right reservoirs, respectively. $\Sigma_{P_n}(\omega)$ is the self-energy of the $n$-th probe. The  hybridization matrix is obtained as $\Gamma_{\alpha}(\omega)=-2 \, {\rm Im}[\Sigma_{\alpha}(\omega)]$ where $\alpha=L,R, P$. For simplicity, we approximate the self energy terms by the wide-band limit, i.e., $\Sigma_{L}(\omega)|_{11}=-i \tau/2$, $\Sigma_{R}(\omega)|_{NN}=-i \tau/2$, and $\Sigma_{P_n}|_{nn}=-i \tau_{p}/2$.

Following the B\"uttiker probe technique, different types of incoherent effects can be realized. For example, at a given voltage bias between the left and the right reservoirs and uniform finite temperature at all terminals, incoherent elastic scattering processes can be implemented by demanding frequency resolved zero charge and energy current between each probe and the system. Such probes are often refereed to as dephasing probes. On the other hand, incoherent inelastic processes can be implemented via the so-called voltage probe technique where the net charge current flowing between each probe and the system vanishes whereas the energy current remains finite.

Since in our work, we focus on electronic conductance, we consider a linear response regime by assuming small chemical potential difference between the left and the right reservoir. The chemical potential for the probes are determined by the zero electronic current conditions. We expand the Fermi-distribution functions in Eq.~(\ref{probe_current}) as 
\begin{align}
\label{linear_response}
f_{\nu} (\omega)=f_{\rm eq}(\omega)-\frac{\partial f_{\rm eq}(\omega)}{\partial \omega}\Big|_{\epsilon_F} (\mu_\nu - \epsilon_F) \nonumber \\
f_{\alpha} (\omega)=f_{\rm eq}(\omega)-\frac{\partial f_{\rm eq}(\omega)}{\partial \omega}\Big|_{\epsilon_F} (\mu_{\alpha} - \epsilon_F),
\end{align}
where $\epsilon_F$ is the chosen chemical potential at equilibrium. Now putting Eq.~(\ref{linear_response}) in Eq.~(\ref{probe_current}), we get an expression for the electronic current flowing out of the $n$-th probe as
\begin{align}
\label{probe_current1}
\mathcal{I}_n =& \sum_{\alpha} \int_{-\infty}^{\infty} \mathcal{T}_{n \alpha}(\omega) \Big(-\frac{\partial f_{\rm eq} (\omega)}{\partial \omega}\Big)(\mu_n - \mu_{\alpha}) \, d\omega \nonumber \\  
=& \Bigg[\mu_n \sum_{\alpha} \int_{-\infty}^{\infty} \mathcal{T}_{n \alpha}(\omega) \Big(-\frac{\partial f_{\rm eq} (\omega)}{\partial \omega}\Big) d\omega \nonumber \\ 
&- \sum_{n^{\prime}=1}^{N}\mu_{n^{\prime}} \int_{-\infty}^{\infty} \mathcal{T}_{n n^{\prime}}(\omega) \Big(-\frac{\partial f_{\rm eq} (\omega)}{\partial \omega}\Big) d\omega \nonumber \\ 
& - \mu_L \int_{-\infty}^{\infty} \mathcal{T}_{n L}(\omega) \Big(-\frac{\partial f_{\rm eq} (\omega)}{\partial \omega}\Big) d\omega \nonumber \\ 
&-\mu_R \int_{-\infty}^{\infty} \mathcal{T}_{n R}(\omega) \Big(-\frac{\partial f_{\rm eq} (\omega)}{\partial \omega}\Big) \, d\omega \Bigg].
\end{align}
The voltage probe condition demands $\mathcal{I}_n=0$ for each $n=1, 2, \cdots N$ leading to $N$ linear equations which after solving yields unique probe chemical potentials. This analysis further simplifies Eq.~(\ref{probe_current1}) in the zero temperature limit ($\beta \to \infty$) to yield
\begin{align}
\label{probe_current2}
\mu_n \sum_{\alpha} \mathcal{T}_{n \alpha}(\epsilon_F) - \sum_{n^{\prime}=1}^{N}\mu_{n^{\prime}} \mathcal{T}_{n n^{\prime}}(\epsilon_F) \\ \nonumber
= \mu_L \mathcal{T}_{nL}(\epsilon_F)  
+\mu_R  \mathcal{T}_{nR}(\epsilon_F).
\end{align}
A more compact way of expressing the above equation is
\begin{align}
\label{probe_current3}
(\mu_n-\mu_R) \sum_{\alpha} \mathcal{T}_{n \alpha}(\epsilon_F)
- \sum_{n^{\prime}=1}^{N}(\mu_{n^{\prime}}-\mu_R)  \mathcal{T}_{n n^{\prime}}(\epsilon_F) \nonumber \\ 
= (\mu_L-\mu_R) \mathcal{T}_{n L}(\epsilon_F).
\end{align}
From Eq.~(\ref{probe_current3}) putting the transmission probabilities explicitly in terms of NEGF, we can write down the solution for chemical potential of each probe $n$ as,
\begin{align}
\label{probesolution}
\mu_n = \mu_R +  \tau \sum_{j=1}^N {\cal W}_{nj}^{-1}(\epsilon_F) \left|G_{j1} (\epsilon_F)\right|^2  (\mu_L - \mu_R) \\ \nonumber
~\forall \, n=1,2, 3 \ldots N
\end{align}


 Here,
the elements of the  $N\times N$ matrix $\mathcal{W}(\epsilon_F)$ are
\begin{align}
\label{W}
\mathcal{W}_{nj} &= -\tau_p |G_{nj}|^2, \quad \forall \, n\neq j \nonumber \\
\mathcal{W}_{nn} & =\tau  (|G_{n1}|^2+|G_{nN}|^2)+\tau_p  \sum_{j\neq n}|G_{nj}|^2,
\end{align}
We consider $\mu_R=\epsilon_F$, $\mu_L=\epsilon_F+ \delta \mu$ and $\mu_n=\epsilon_F+\delta \mu_n$.
By determining the probe chemical potentials $\mu_n$ using Eq.~(\ref{probesolution}), one can find the two-terminal electronic conductance, defined as, $\mathcal{G}(\epsilon_F) = \mathcal{I}_L/\delta \mu$ where $\mathcal{I}_L$ is given in Eq.~(\ref{probe_current}). The conductance is given by \cite{Pastawski},
\begin{align}
\label{conductance}
\mathcal{G}(\epsilon_F)  &= \tau^2  \left| G_{1N}(\epsilon_F) \right|^2 \nonumber \\
& +   \tau^2\tau_p \sum_{n,j=1}^{N}\left|G_{Nn} (\epsilon_F)\right|^2 
\mathcal{W}^{-1}_{nj}(\epsilon_F) \left|G_{j1} (\epsilon_F)\right|^2 \hspace*{-2pt}
\end{align}
which was given in the main text. 
\section{Details about perturbative expansion of $ \mathcal{W}$ upto ${\cal O}(\tau_p)$}
\begin{figure}
\includegraphics[width=1.0 \columnwidth]{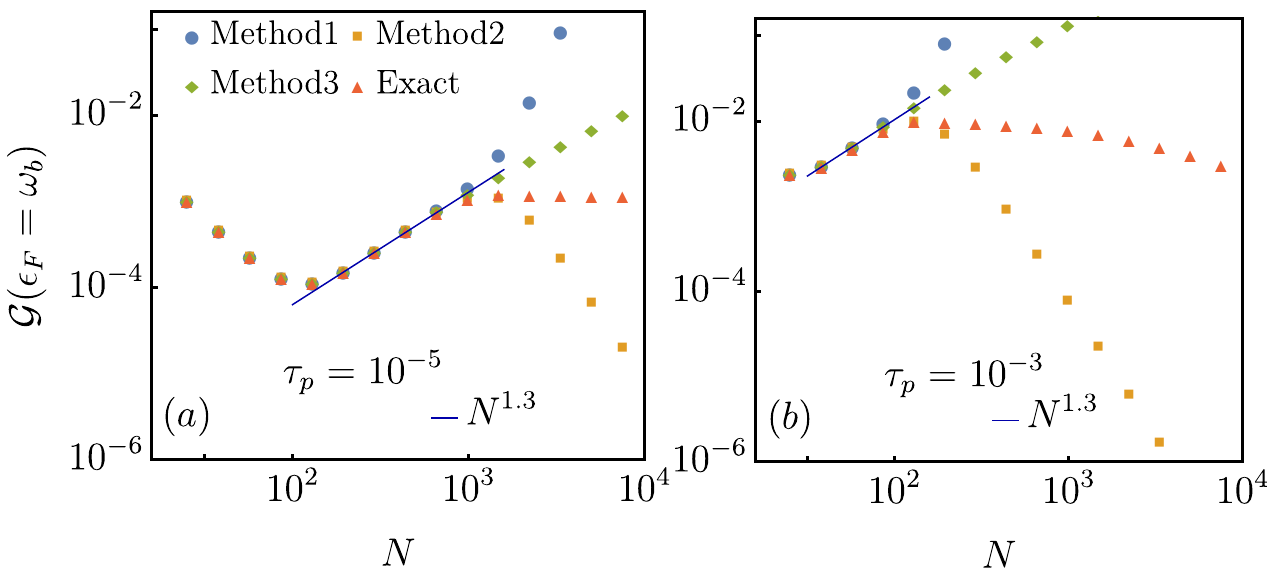} 
\caption{(Color online): Plot for  conductance in presence of B\"uttiker probes at the band edge ($\epsilon_F=\omega_b$). The exact result is compared with three different perturbative methods, as mentioned in the text. The probe couplings are (a) $\tau_p=10^{-5}$ and (b) $\tau_p=10^{-3}$. The superballistic regime is perfectly captured by all the three methods. The deviation from the exact result occur just after the termination of the superballistic regime.}
\label{numerical_check1} 
\end{figure}

In this section, we provide details about the perturbative expansion of $ \mathcal{W}$ upto ${\cal O}(\tau_p)$. We write $ \mathcal{W}$ as, 
\begin{equation}
    \mathcal{W}_{lm} =  \mathcal{W}^{(1)}_{lm} + \tau_p \mathcal{W}^{(2)}_{lm} + \mathcal{O}(\tau_p^2).
    \label{w-perturbative}
\end{equation}
We now get the components of  $\mathcal{W}$ by performing the perturbative expansion in NEGF, as given in Eq.~(\ref{Gr}).
Upto ${\cal O}(\tau_p)$ we get
\begin{align}
G(\omega)&=\Big[\omega \mathbb{I}-H_S-\Sigma_L-\Sigma_R-\sum_{n=1}^{N} \Sigma_{P_n}\Big]^{-1}  \nonumber \\
&=\Big[(G^{0})^{-1}-\sum_{n=1}^{N} \Sigma_{P_n}\Big]^{-1} \nonumber \\
&=\Big[(G^{0})^{-1}+\frac{i \tau_p}{2} \mathbb{I}\Big]^{-1} \nonumber \\
&= G^{0} - \frac{i \tau_p}{2} G^{0} G^{0} + \mathcal{O}(\tau_p^2)
\label{approxgreen}
\end{align} 
where recall that $G^{0}$ is the retarded Green's function in absence of the probes. We have suppressed the argument $\omega$ for sake of brevity. As we are interested upto $\tau_p$ terms, following Eq.~(\ref{approxgreen}), the off-diagonal term $(l \neq m)$  of $\mathcal{W}$  can be simply replaced by  $-\tau_p |G^{0}_{lm}|^2$. Similarly, the second term in the the diagonal piece $(l = m)$  of $\mathcal{W}$, given in Eq.~(\ref{W}), can be replaced by $\tau_p  \sum_{\alpha=1}^{N}|G^{0}_{l\alpha}|^2$. The term proportional to $\tau $ in Eq.~(\ref{W}) needs to be carefully analyzed. For this purpose, an expansion in $\tau_p$ for $|G_{l1}|^2$ is carried out and is given by,
\begin{align}
 |G_{l1}|^2 &=
 |G^{0}_{l1}|^2 \nonumber \\
 &- \frac{i \tau_p}{2}\Big[G^{0a}_{1l}(G^{0} G^{0})_{l1}\!-\! G^{0}_{l1}(G^{0a} G^{0a})_{1l}\Big]  +  \mathcal{O}(\tau_p^2),
\end{align}
where $G^{0a}$ is the advanced Green's function without the probes. In a similar fashion,
\begin{align}
 |G_{lN}|^2 & = |G^{0}_{lN}|^2 \nonumber \\
& \!-\!\! \frac{i \tau_p}{2}\Big[G^{0a}_{Nl}(G^{0} G^{0})_{lN}\!-\! G^{0}_{lN}(G^{0a} G^{0a})_{Nl} \Big]\!+\! \mathcal{O}(\tau_p^2).
\end{align}
As a result, Eq.~(\ref{W}), up to $\mathcal{O}(\tau_p)$, is given as,
\begin{equation}
\mathcal{W}^{(1)}_{lm} = 
\begin{cases}
    0,& \forall \, l\neq m\\
    \tau \Big(|G_{l1}^0|^2+|G_{lN}^0|^2\Big),& \forall\,  l=m 
\end{cases}
\label{wmatrixelement3}
\end{equation}
and 
\begin{widetext}
\begin{equation}
\mathcal{W}^{(2)}_{lm} = 
\begin{cases}
    -|G_{lm}^0|^2,& \forall l\neq m\\
     \sum_{\alpha=1}^{N}|G_{l\alpha}^0|^2 +\frac{i \tau}{ 2} \Big[G^{0}_{lN}(G^{0a} G^{0a})_{Nl}+G^{0}_{l1}(G^{0a} G^{0a})_{1l}-G^{0a}_{Nl}(G^{0} G^{0})_{lN}-G^{0a}_{1l}(G^{0} G^{0})_{l1}\Big] , &\forall l=m 
\label{wmatrixelement4}
\end{cases}
\end{equation}
\end{widetext}
To check the validity of our approximation upto $\mathcal{O}(\tau_p)$ in $\mathcal {W}$, we have computed the two-terminal conductance using three different methods:

\noindent {\it Method 1}: We keep terms up to $\mathcal{O}(\tau_p)$ in both $\mathcal{W}$ and in the NEGF which appears in the second term of two-terminal conductance in presence of probes [see Eq.~(\ref{conductance})]. NEGF's up to $\mathcal{O}(\tau_p)$ are
\begin{align}
    |G_{Nn}|^2 &= |G_{Nn}^0|^2 \nonumber \\
    & +i \frac{ \tau_P}{2}\Big[G^{0}_{Nn} (G^{0a}G^{0a})_{nN}-G^{0a}_{nN} (G^{0}G^{0})_{Nn}\Big], \nonumber \\
   |G_{j1}|^2 &=|G_{j1}^0|^2 \nonumber \\
    &+i \frac{ \tau_P}{2} \Big[G^{0}_{j1} (G^{0a}G^{0a})_{1j}-G^{0a}_{1j} (G^{0}G^{0})_{j1}\Big].
    \end{align}
{\it Method 2:} We keep terms up to $\mathcal{O}(\tau_p)$ in $\mathcal{W}$ but  for the NEGFs we keep $\tau_p$ independent terms  which are given by, 
\begin{align}
|G_{Nn}|^2 & =|G_{Nn}^0|^2, \nonumber \\
|G_{j1}|^2 & =|G_{j1}^0|^2.
\end{align}
{\it Method 3:} We approximate exact expression for $\mathcal{W}$ [see Eq.~(\ref{W})] by replacing the NEGF $G$ by $G^{0}$ and also for the NEGF's in Eq.~(\ref{conductance}). As a result the NEGFs have the form same as in method 2, whereas 
the form of $\mathcal{W}$ is given by,
    \begin{equation}
\mathcal{W}_{lm} = 
\begin{cases}
    -\tau_p |G^0_{lm}|^2, \quad \forall \,\, l\neq m \\
    \tau  (|G^0_{l1}|^2+|G^0_{lN}|^2)+\tau_p  \sum_{\alpha=1}^{N}|G^0_{l\alpha}|^2, \, \forall \, l=m.
\end{cases}
\label{wmatrixelement2}
\end{equation}
Remarkably, all the above three methods perfectly capture the superballistic transport regime for conductance with system length $N$. In Fig.~\ref{numerical_check1}, we display the conductance at the band edge obtained from the three methods. We see a clear evidence of the SB transport regime captured by all the methods and matches perfectly with the exact numerics. However beyond the SB regime, all the above mentioned perturbative approximations fail to capture the exact trend in conductance.

\begin{figure}
\includegraphics[width=1.0 \columnwidth]{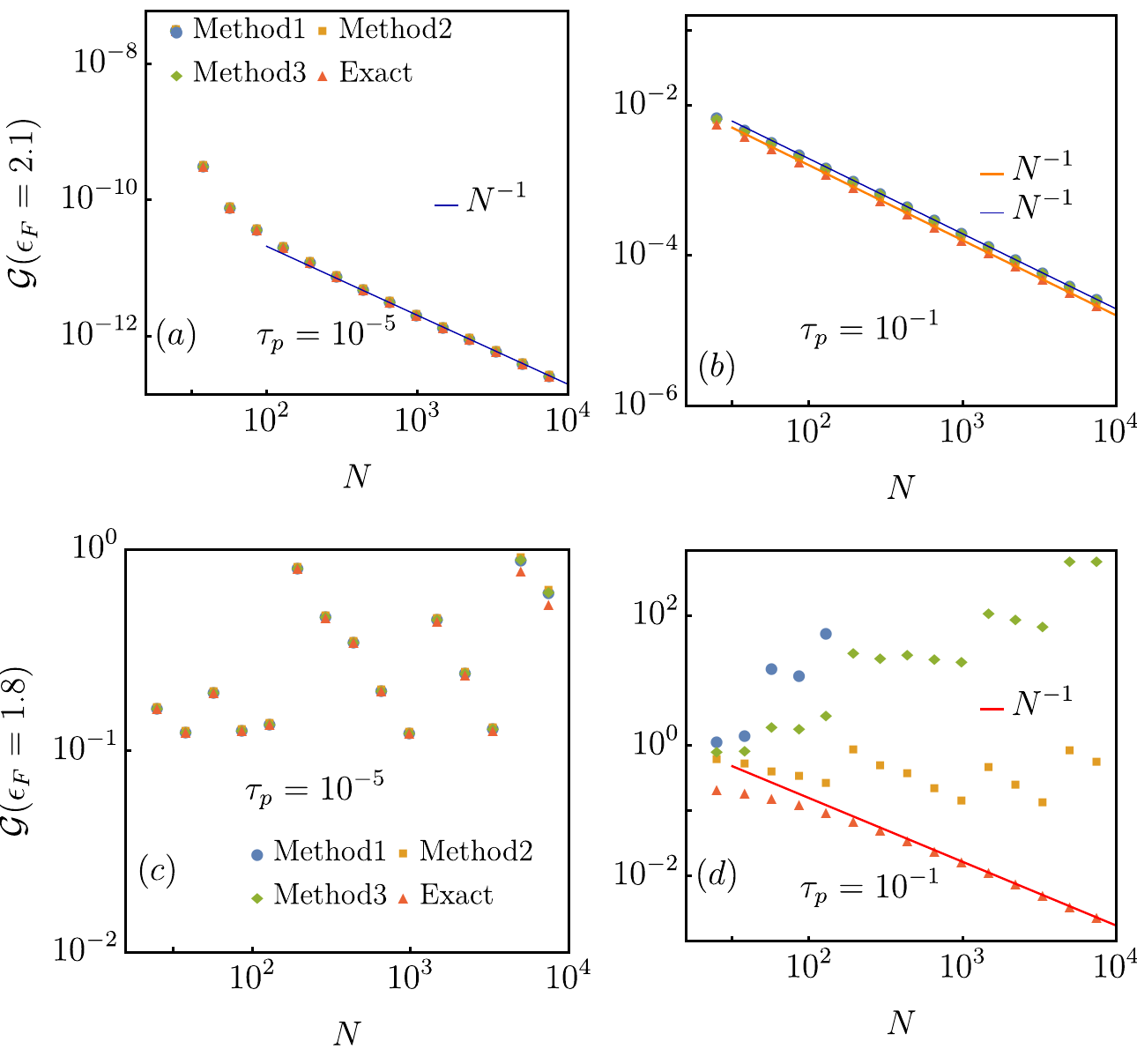} 
\caption{(Color online): Plot for conductance outside ($\epsilon_F=2.1$) [(a) and (b)] and inside ($\epsilon_F=1.8$) of the band edge [(c) and (d)]  for two different probe couplings $\tau_p=10^{-5}$ and $\tau_p=10^{-1}$, respectively. For weak probe coupling [(a) and (c)], all the three methods agree irrespective of whether the $\epsilon_F$ is inside or outside the band edge. In contrast, for strong probe coupling [(b) and (d)], the perturbative results differ from the exact numerics. Interestingly, for $\epsilon_F=2.1$, (b) all three methods correctly capture the diffusive behaviour. In contrast, for $\epsilon_F=1.8$, at strong coupling (d) all the three methods fail to capture the diffusive behaviour in conductance.}

\label{numerical_check2} 
\end{figure}

\begin{figure}
\includegraphics[width=1.0 \columnwidth]{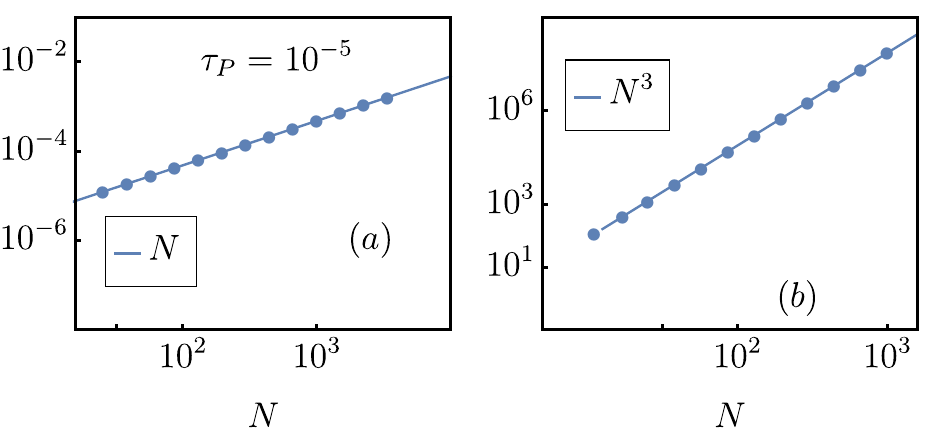} 
\caption{(Color online): Plot for (a) Eq.~(\ref{second-cond}), and (b) $\sum_{n=1}^{N} f(n)$ as a function of $N$ at the band edge.}
\label{numerical_check3} 
\end{figure}

\section{Validity of different perturbative methods outside and within the band edge}
In this section, we consider the scenario when the chemical potential is located either outside or within the band edge. We show the comparison between the results obtained following the different perturbative methods and the exact numerics for the electronic conductance for both weak and strong probe coupling strengths. Fig.~\ref{numerical_check2} (a) and (b) shows that all the three perturbative methods perfectly capture the crossover from exponentially suppressed to diffusive transport when the chemical potential is lying outside the band edge ($\epsilon_F=2.1$). In the other scenario, i.e, within the band edge ($\epsilon_F=1.8$) [Fig.~\ref{numerical_check2} (c) and (d)] results for conductance perfectly match with exact numerics for weak coupling with all three methods. However, for strong coupling all three perturbative methods fail to capture the diffusive trend predicted by exact calculations. This analysis shows that the specific perturbative methods studied in this work are not suitable for capturing ballistic to diffusive crossover. Remarkably, at and outside the band edge, the perturbative methods are highly efficient in predicting the crossover in transport features. 

\section{Linear $N$ scaling in conductance -- understanding the onset of Superballistic transport}
In this section, we are interested to find the scaling of conductance with system size. We focus on the conductance expression that is valid upto the linear order in $\tau_p$, given as 
\begin{align}
 \label{Teff1-SB}
 \mathcal{G}(\epsilon_F)&=\tau^2 |G_{1N}^0(\epsilon_F)|^2 \\ \nonumber
& +\tau \tau_p \sum_{n=1}^{N} \! \frac{|G_{Nn}^0(\epsilon_F)|^2  |G_{n1}^0(\epsilon_F)|^2}{|G_{n1}^0(\epsilon_F)|^2+|G_{Nn}^0(\epsilon_F)|^2} 
+ \mathcal{O}(\tau_p^2). \nonumber 
\end{align}
Note that, at the band edge, the first term provides a subdiffusive $1/N^2$ scaling in conductance \cite{saha2022universal}. We therefore focus on the second term 
\begin{equation}
\tau \tau_p \sum_{n=1}^{N} \! \frac{|G_{Nn}^0(\epsilon_F)|^2  |G_{n1}^0(\epsilon_F)|^2}{|G_{n1}^0(\epsilon_F)|^2+|G_{Nn}^0(\epsilon_F)|^2}
\label{second-cond}
\end{equation}
and analyze the system size $N$ dependence. Recall that, 
$G^0$ is the NEGF matrix for the system in absence of probes. As the single-particle Hamiltonian of the system is tridiagonal, the NEGF in this case can be solved exactly, and given as,
\begin{equation}
G^0_{\ell j}(\omega)=(-1)^{\ell +j} \frac{\Delta_{1,\ell-1}(\omega) \Delta_{N-j,N}(\omega)}{\Delta_{1,N}(\omega)},
\label{det}
\end{equation}
where $\Delta_{i,j}(\omega)$ is the determinant of a part of the matrix $\mathcal{M}(\omega)$, given as
\begin{align}
\mathcal{M}(\omega)= \begin{pmatrix}
\omega -\Sigma_L|_{1,1} & 1 & 0 & 0 & \ldots \\
1& \omega & 1 & 0 & \ldots \\
0 & 1 & \omega &1 & \ldots \\
\ldots & \ldots & \ldots & \ldots \\
0 & 0 & 0 & 1 & \omega -\Sigma_R|_{N,N}
\end{pmatrix}.
\end{align}
The part of the matrix of $\mathcal{M}(\omega)$ is constructed by starting from 
$i$-th row and column and ending with $j$-th row and column.  Note that, albeit we use the same notation $\Delta_{i,j}(\omega)$ as in main text, here we are using it in the context of Green's function in absence of probes. The determinants $\Delta_{1,N}(\omega)$, $\Delta_{1,\ell-1}(\omega)$ and $\Delta_{N-p,N}(\omega)$ that appears in Eq.~(\ref{det}) can be computed following the transfer matrix approach \cite{saha2022universal}. The equations which solve the determinants are following,
\begin{eqnarray}
\begin{pmatrix}
\Delta_{1,N}  \\
\Delta_{1,N-1}
\end{pmatrix} &=& \begin{pmatrix}
1 & -\Sigma_L|_{1,1}  \\
0 & 1
\end{pmatrix} \begin{pmatrix}
\omega & -1 \\
1 & 0
\end{pmatrix}^{N}  \begin{pmatrix}
 1 \\
 \Sigma_R|_{N,N}
\end{pmatrix}, \nonumber \\
\begin{pmatrix}
\Delta_{1,\ell-1}  \\
\Delta_{1,\ell-2} 
\end{pmatrix} &=& \begin{pmatrix}
1 & -\Sigma_L|_{1,1}  \\
0 & 1
\end{pmatrix} \begin{pmatrix}
\omega & -1 \\
1 & 0
\end{pmatrix}^{\ell-1}  \begin{pmatrix}
 1 \\
 0
\end{pmatrix},\\
\begin{pmatrix}
\Delta_{N-p,N} \\
\Delta_{N-p-1,N}
\end{pmatrix} &=&  \begin{pmatrix}
\omega & -1 \\
1 & 0
\end{pmatrix}^{N-p}  \begin{pmatrix}
 1 \\
 \Sigma_R|_{N,N}
\end{pmatrix}. \nonumber 
\label{determinants}
\end{eqnarray}
At the band edges $\omega_b=\pm 2$, the transfer matrix is non-diagonalizable and has exceptional points. However, one can write the matrix in  the Jordon-normal form $\begin{pmatrix}
1 & 1 \\
0 & 1
\end{pmatrix}$ by using a transformation $S=\begin{pmatrix}
i & 0 \\
i & -i
\end{pmatrix}$. With these in hand, we get the specific forms of the determinants, given as 
\begin{eqnarray}
\Delta_{1,N}(\omega_b)&=&\frac{3N+5}{4}+ i\, N, \nonumber \\ 
\Delta_{1,\ell-1}(\omega_b) &=&l + i \, \Big(\frac{\ell-1}{2}\Big), \\ 
\Delta_{N-p,N}(\omega_b) &=& \Big(N-p+1\Big) + i \Big(\frac{N-p}{2}\Big) \nonumber .
\end{eqnarray}
Finally using Eq.~(\ref{second-cond}), we get
\begin{align}
&\tau \tau_p \sum_{n=1}^{N} \! \frac{|G_{Nn}^0(\omega_b)|^2  |G_{n1}^0(\omega_b)|^2}{|G_{n1}^0(\omega_b)|^2+|G_{Nn}^0(\omega_b)|^2}  \nonumber \\
&\qquad \quad \qquad=\frac{\tau \tau_p}{|\Delta_{1,N}(\omega_b)|^2}\sum_{n=1}^{N} f(n) \propto N,
\end{align}
where $f(n)$ is given by
\begin{equation}
    f(n)=\frac{|\Delta_{1,n-1}(\omega_b)|^2 |\Delta_{N-n,N}(\omega_b)|^2}{|\Delta_{1,n-1}(\omega_b)|^2 + |\Delta_{N-n,N}(\omega_b)|^2}.
    \label{feq}
\end{equation}
We find that 
\begin{equation}
    \sum_{n=1}^{N} f(n) \propto N^3
\end{equation} and is clearly shown in Fig.~\ref{numerical_check3}(b). Also from Eq.~(\ref{determinants}) we find 
\begin{equation}
    |\Delta_{1,N}(\omega_b)|^2 \propto N^2.
\end{equation} 
As a result, upto the linear order in $\tau_p$, the second term in the conductance in Eq.~(\ref{Teff1-SB}) scales linearly with $N$ at the band-edge, as shown in Fig.~\ref{numerical_check3}(a). This term is therefore responsible for the onset of the superballistic transport regime. 

\section{Particle density in the absence and the presence of probes}
\begin{figure}
\includegraphics[width=1.0 \columnwidth]{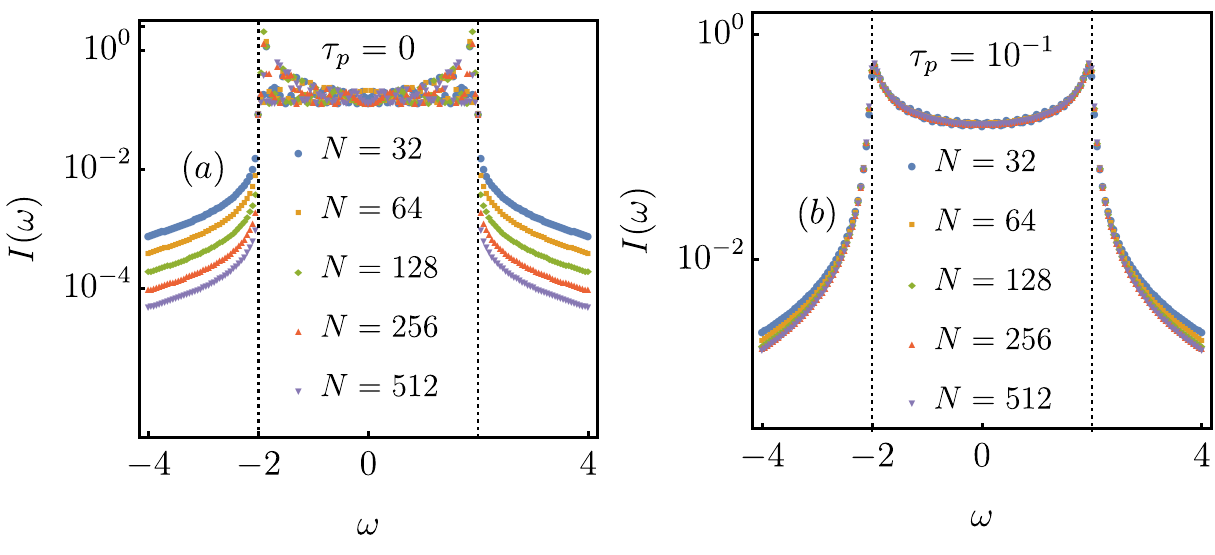} 
\caption{(Color online):  We plot integrand $I(\omega)$, defined in Eq.~(\ref{integrand-I}), of the particle density as a function of $\omega$ for different $N$ with weak to strong probe couplings. }
\label{particle_density} 
\end{figure}
In this section, we show that the effect of having voltage probes is very different from the effect of having many-body interactions in the system, although both are sources of inelastic scattering. To this end, we calculate the particle density of the system which is defined by,
\begin{equation}
    \gamma=\frac{1}{N} \sum\limits_{\ell=1}^N \langle c_{\ell}^{\dagger}c_{\ell}\rangle.
    \label{particle-density}
    \end{equation}
At equilibrium, the occupation at the $\ell$th site in presence of probes can be written as,
\begin{align}
\langle c_{\ell}^{\dagger}c_{\ell}\rangle&=\tau \int\limits_{-\infty}^{\epsilon_F}(|G_{1\ell}(\omega)|^2+|G_{N\ell}(\omega)|^2) \frac{d\omega}{2\pi}  \\ \nonumber 
& + \tau_p \int\limits_{-\infty}^{\epsilon_F}\sum\limits_{\alpha=1}^N|G_{\alpha\ell}(\omega)|^2 \frac{d\omega}{2\pi}
\end{align} 
where we recall that $\epsilon_F$ is the Fermi energy. Thus, the full integrand in the particle density is 
\begin{eqnarray}
    I(\omega)&=&\frac{1}{2\pi N}\sum\limits_{\ell=1}^N\Big[\tau (|G_{1\ell}(\omega)|^2+|G_{N\ell}(\omega)|^2)\nonumber \\
    &+& \tau_p \sum\limits_{\alpha=1}^N|G_{\alpha\ell}(\omega)|^2 \Big]
    \label{integrand-I}
    \end{eqnarray}
    
and therefore 
\begin{equation}
\gamma=\int\limits_{-\infty}^{\epsilon_F} I(\omega) d\omega.
\label{integrand}
\end{equation}
To conclude about the $N$ dependence of $\gamma$, we plot $I(\omega)$ for different system lengths $N$. In absence of probes, i.e, $\tau_p=0$, we see in Fig.~\ref{particle_density}(a) that $I(\omega)$ decreases with $N$ for $\omega\leq-2$, i.e, $\omega$ below lower band-edge. This means particle density at lower band-edge, i.e, $\gamma$ for $\epsilon_F=-2$, decreases with $N$. Thus, as system-size increases, the particle density goes to zero. As a consequence of vanishing particle density at large system length, the transport mechanism is governed by the single-particle sector. So, even if number conserving many-body interactions are switched on within the lattice chain, it will have a negligible effect. Consequently, the sub-diffusive behavior, that is seen at the band edges in absence of many-body interactions, survives even in presence of interactions.

In Fig.~\ref{particle_density}(b) we plot $I(\omega)$ for different system lengths $N$ for $\tau_p=0.1$. We clearly see that on increasing system-size, $I(\omega)$ collapses to a single curve. This means particle density at lower band-edge, i.e, $\gamma$ for $\epsilon_F=-2$ saturates with increasing $N$. Thus, contrary to the situation in absence of probes, in presence of probes there is finite particle density at large $N$ for $\epsilon_F=-2$. So, even though having many-body interactions (in the system) or the presence of probes are sources of inelastic scattering, they have remarkably different effect at the lower band-edge. Similar conclusion can be reached for the upper band-edge, $\epsilon_F=2$, when considering holes instead of particles.

\begin{figure}
\includegraphics[width=1.0 \columnwidth]{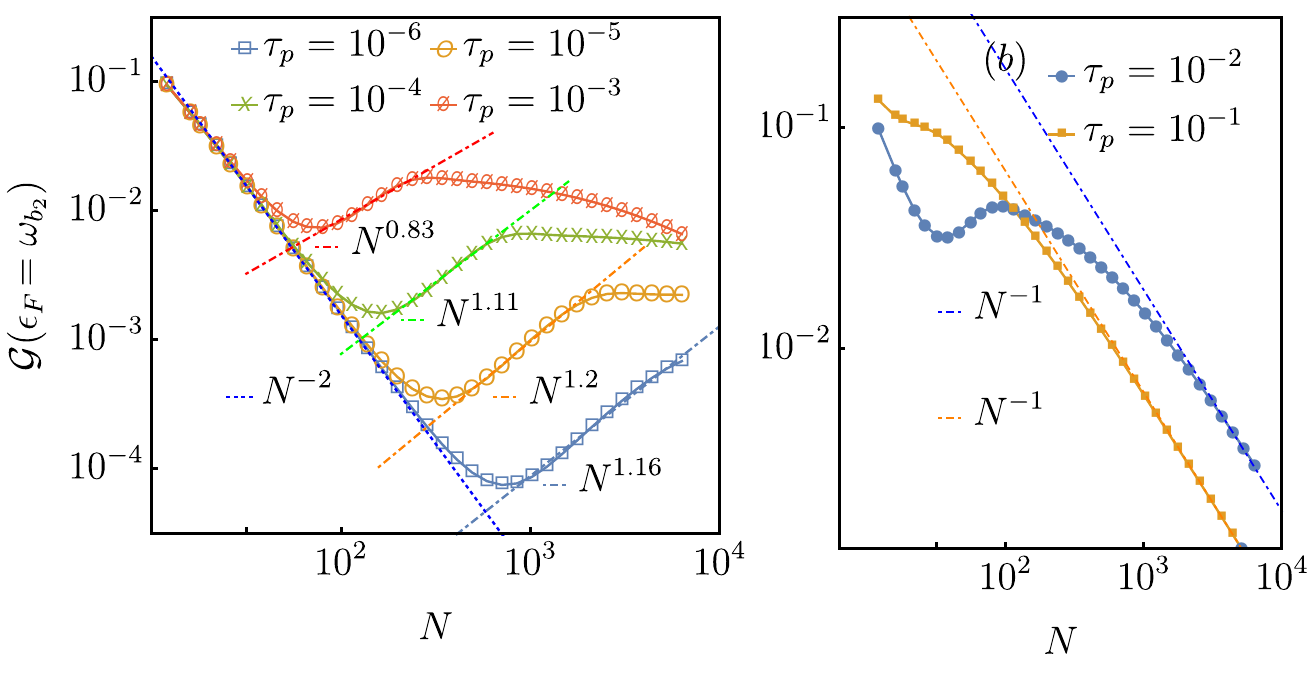} 
\caption{(Color online):(a) Behavior of conductance at one of the band-edges $\epsilon_F=\omega_{b_2}=-W$ of a two-band model given in Eq.~(\ref{onsite}) vs system size for small $\tau_p$. The sub-diffusive to superballistic crossover is clearly seen. (b) Plots of conductance at larger values of $\tau_p$ which captures the eventual crossover to conventional diffusive regime at large $N$. All the other band-edges show similar behavior. Here we choose $W=0.5$.}
\label{two_band1} 
\end{figure}

\section{Eigenvalues of $\mathbf{T}(\omega)$ and NEGF in presence of probes}
The matrix $\mathbf{T}(\omega)$ of a normal tight-binding model in presence of probes is given by,
\begin{align}
\mathbf{T}(\omega)=\mathbf{T}^0(\omega)+\frac{i \tau_p}{4} (\mathbb{I}_2+\sigma_z)    
\end{align}
Here $\mathbf{T}^0(\omega)=\frac{\omega}{2}(\mathbb{I}_2+\sigma_z)-i\sigma_y$. Thus, $\mathbf{T}(\omega)=(\frac{\omega}{2}+ \frac{i \tau_p}{4})(\mathbb{I}_2+\sigma_z)-i\sigma_y$. The eigenvalues of $\mathbf{T}(\omega)$ is, 
\begin{align}
\lambda_{\pm}=\frac{1}{2}\left[\omega+\frac{i\tau_p}{2} \pm \sqrt{\left(\omega+\frac{i\tau_p}{2}\right)^2-4} \right]
\label{eigenval11}
\end{align}
This can be written as, $\lambda_{\pm}=e^{\pm (\kappa_1+ i\kappa_2)}$, with $\kappa_1, \kappa_2 \geq 0$ and $\kappa_1=\mathrm{log}|\lambda_{+}|$.
\subsection{NEGF in presence of probes for all $\omega$}
It can be checked that with $\tau_p>0$ for all $\omega$, in Eq.~(\ref{eigenval11}), absolute value of one of the eigenvalues $|\lambda_{+}|$ is greater than 1, while the other $|\lambda_{-}|$ is less than 1. It ensures that $\kappa_1>0$. For the NEGF, we know that
\begin{equation}
|G_{\ell j}(\omega)|^2 =  \frac{|\Delta_{1,\ell-1}(\omega)|^2 |\Delta_{N-j,N}(\omega)|^2}{|\Delta_{1,N}(\omega)|^2}.
\label{negf}
\end{equation}
Since the determinant $\Delta_{i,j}$ is related to $\mathbf{T}(\omega)$, calculating it requires calculating powers of $\mathbf{T}(\omega)$. The value of $\Delta_{i,j}$ is therefore dominated by the eigenvalue with larger magnitude, i.e., $\lambda_+$. So, we find that, $|\Delta_{1,\ell-1}(\omega)|^2 \sim e^{2\kappa_1 \ell}$, $|\Delta_{N-j,N}(\omega)|^2 \sim e^{2\kappa_1 (N-j)}$ and $|\Delta_{1,N}(\omega)|^2 \sim e^{2\kappa_1 N}$. Putting all these values in Eq.~(\ref{negf}), we get $|G_{\ell j}(\omega)|^2\sim e^{-2\kappa_1 |l-j|}$. Thus, if we write $|G_{\ell j}(\omega)|^2\sim e^{-|l-j|/\xi}$, ($\xi$ is the localization length), then $\xi^{-1}=2 \kappa_1$.

\subsection{Small $\tau_p$ behaviour of $|\lambda_{\pm}|$ at the band edges}
The eigenvalues of $\mathbf{T}(\omega)$ at one of the band edges $\omega_b=2$ reduces to,
\begin{align}
\lambda_{\pm}&=\frac{1}{2}\left[2+\frac{i\tau_p}{2} \pm \sqrt{\left(-\frac{\tau_p^2}{4}+2i\tau_p\right)} \right] \\ \nonumber
&=1+\frac{i\tau_p}{4}\pm \sqrt{\tau_p}\sqrt{-\frac{\tau_p}{16}+\frac{i}{2}}.
\end{align}
Defining $\theta=\tan^{-1}(8/\tau_p)$, we can write the above equation as,
\begin{align}
\label{eigenval}
\lambda_{\pm}&=1+\frac{i\tau_p}{4}\pm \sqrt{\tau_p} \, \exp(-i \theta/2)~~~~~~ \\ \nonumber
&=1\pm \sqrt{\tau_p}\cos(\theta/2)+\frac{i\tau_p}{4}\mp i\sqrt{\tau_p}\sin(\theta/2),
\end{align}
where  $\cos(\theta)=\frac{\tau_p}{\sqrt{64+\tau_p^2}}$ and $\sin(\theta)=\frac{8}{\sqrt{64+\tau_p^2}}$. 
Taking the absolute value of Eq.~(\ref{eigenval}) we obtain
\begin{align}
|\lambda_{\pm}|&=\sqrt{\Big(1\pm \sqrt{\tau_p}\cos(\theta/2)\Big)^2+\Big(\frac{\tau_p}{4}\mp \sqrt{\tau_p}\sin(\theta/2)\Big)^2} \nonumber \\ 
&=\sqrt{1+\tau_p+\frac{\tau_p^2}{16}
\pm 2\sqrt{\tau_p}\cos(\theta/2)\mp  \frac{\tau_p^{3/2}}{2} \sin(\theta/2)} \nonumber \\ 
&\approx \sqrt{1\pm 2 \sqrt{\tau_p}} ~~~~~\mathrm{small} ~\tau_p \nonumber \\
&\approx 1\pm \sqrt{\tau_p}.   
\end{align}
Exactly same result can be obtained for the lower band edge, $\omega_b=-2$ as well. 

\section{Superballistic scaling of conductance at the band edges of two-band model}

\begin{figure}
\includegraphics[width=1.0 \columnwidth]{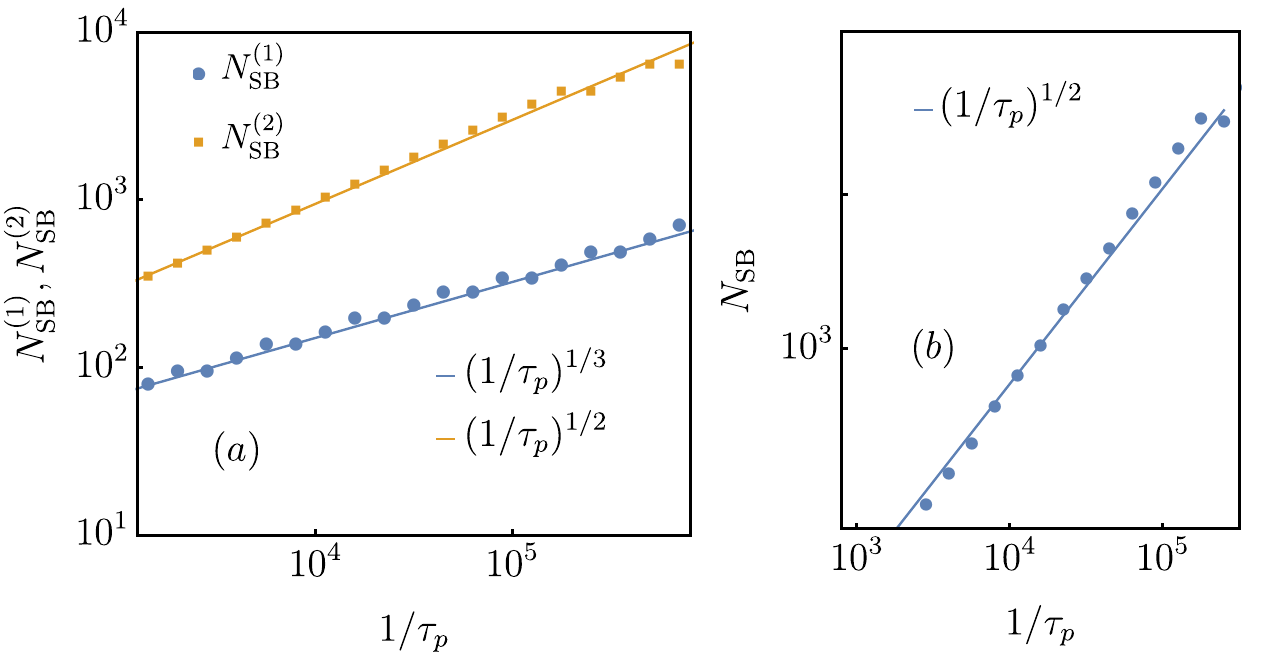} 
\caption{(Color online): (a) The scaling of the system-size corresponding to the start (end) of the superballistic scaling regime $N_{\rm SB}^{(1)}$ ($N_{\rm SB}^{(2)}$) with $\tau_p^{-1}$ for the two-band model given in Eq.~(\ref{onsite}).  (b) The scaling of the size of the superballistic scaling regime $N_{\rm SB}=N_{\rm SB}^{(2)}-N_{\rm SB}^{(1)}$ with $1/\tau_p$. This clearly shows that the superballistic scaling regime expands on decreasing $\tau_p$. }
\label{two_band2} 
\end{figure}
In this section, we discuss the situation when the tight-binding lattice supports two bands. Let us consider the periodic onsite potential term as 
\begin{equation}
V= \sum_{i=1}^N \epsilon_i c_i^{\dagger}c_i\!=\!W \sum_{i=1}^N \cos[2\pi b i] \, c_i^{\dagger}c_i
\label{onsite}
\end{equation} 
with the usual nearest neighbour hopping term discussed in the main text. Here $i$ is the site index, $W$ is the strength of the onsite potential and we choose $b=1/2$ i.e., the two band case. The unit cell transfer matrix without the probes for this case is, 
\begin{equation}
   \mathbf{T}^0(\omega)=\prod\limits_{j=1}^{2}\mathbf{T}^{0(j)}(\omega) 
\end{equation} with 
\begin{equation}
    \mathbf{T}^{0(j)}(\omega)=\frac
{\omega-\epsilon_j}{2}(\mathbb{I}_2 + \sigma_z)-i \sigma_y.
\end{equation} 
Here $\mathbf{T}^{0(j)}(\omega)$ is the transfer matrix for the each site $j$. At the band edges of the two-band case ($\omega=\omega_{b_2}$), \begin{equation}
\mathrm{Tr}\big[\mathbf{T}^{0}(\omega\!=\!\omega_{b_2})]\big]=\pm 2.  
\label{dis-cond}
\end{equation}
The subscript ${b_2}$ in $\omega_{b_2}$ indicates the fact that we are considering the two-band case. Solving the above condition [Eq.~(\ref{dis-cond})], one will get four band edges at the values, $\omega_{b_2}=\pm W, \pm \sqrt{4+ W^2}$. Now, in presence of weak environment effects, we show the superballistic behaviour for one of the band edges $\omega_{b_2}=-W$ of two-band model  in Fig.~\ref{two_band1} (a). With increasing probe strength, the crossover from superballistic to diffusive regime is also visible in Fig.~\ref{two_band1} (b). The superballistic regime also can be expanded like the one-band model which we have shown in Fig.~\ref{two_band2}. These findings hold for any chosen band edge and is generalisable to chains that host more than two bands.

\end{document}